\documentclass{article}
\usepackage[utf8]{inputenc}
\usepackage{geometry}
\usepackage{setspace}
\usepackage{titling}
\usepackage{graphicx} 
\usepackage[backend=biber, style=apa, citetracker=true,doi=false,
  url=false]{biblatex} 
\usepackage{multirow}
\usepackage{adjustbox}
\usepackage{subcaption}
\usepackage[labelsep=period]{caption}
\usepackage{booktabs}
\usepackage{rotating}
\usepackage{enumitem}
\usepackage{graphicx}
\usepackage{indentfirst}
\usepackage{amsmath,amsthm}
\usepackage{thm-restate}
\usepackage{amssymb}
\usepackage{mathtools}
\usepackage{bbm}
\usepackage{bm}
\usepackage{float}
\usepackage[flushleft]{threeparttable}
\usepackage{lipsum}  
\usepackage{authblk}
\usepackage[colorlinks=true, allcolors=blue]{hyperref}
\usepackage{cleveref}

\theoremstyle{plain}
\newtheorem{assumption}{Assumption}[section]

\newtheorem{remark}{Remark}[section]
\newtheorem*{definition}{Definition}

\declaretheorem[name=Corollary]{corollary}

\newcounter{variant}

\DeclareMathOperator*{\rank}{rank} 

\title{Sufficient Statistics for Markovian Feedback Process and Unobserved Heterogeneity in Dynamic Panel Logit Models}
\author{Sukgyu Shin\thanks{Department of Economics, The Ohio State University. Email: \texttt{shin.927@osu.edu}}}

\begin{document}

\maketitle

\begin{abstract}

    In this paper, we examine identification in dynamic panel logit models with state dependence, a first-order Markov feedback process, and individual unobserved heterogeneity by introducing sufficient statistics for the feedback process and the unobserved heterogeneity. If a sequentially exogenous discrete covariate follows a first-order Markov process, identification via conditional likelihood is infeasible regardless of the time period. We also establish the failure of point identification beyond the conditional likelihood framework, which necessitates additional restrictions for identification. We present two assumptions for identification via conditional likelihood, imposed on the feedback process and the initial condition, respectively.

\end{abstract}

\textbf{JEL Codes:} C23, C25

\textbf{Keywords:} Fixed Effect Dynamic Panel Logit, Sequential Exogeneity, Markov Process, Feedback, Sufficient Statistics, Conditional Maximum Likelihood Estimation.

\clearpage
\section{Introduction}
\noindent Many economic problems are inherently dynamic over time in the sense that current choices depend on past choices. Behavioral persistence is a central feature of such settings, which shows the necessity of dynamic discrete choice models. Past choices may also shape subsequent covariates. Even after controlling for additive individual heterogeneity to eliminate potential spurious dynamics, two sources of dynamics remain. One is state dependence, or inertia, whereby individuals are likely to stick to past choices they have made. The other arises when covariates depend on past outcomes, which is referred to as a feedback process. Aside from the state dependence, past outcomes may also affect current outcomes via the feedback process of the covariates.



Despite its wide range of empirical applications, identifying the effects of state dependence and a feedback process in dynamic discrete choice models is difficult. This paper investigates identification and the minimum time periods required for identification in dynamic panel logit models with state dependence, first-order Markov feedback process, and individual unobserved heterogeneity by introducing sufficient statistics for the process and unobserved heterogeneity.

A main challenge in a binary panel logit model arises when there exists individual-specific heterogeneity unobserved from the data. Due to the unobserved heterogeneity, two types of dynamics arise: true dynamics from state dependence and spurious dynamics from unobserved heterogeneity. Distinguishing between these two sources of persistence is nontrivial. If heterogeneity is not factored into the model, the resulting estimates may reflect spurious dynamics driven by the correlation between unobserved heterogeneity and observed covariates, rather than genuine behavioral persistence \parencite{heckman_1978, Heckman1981heterogeneity}. However, the presence of the unobserved heterogeneity in nonlinear settings implies significant identification challenges, unlike in linear panel models, as unobserved heterogeneity is hard to be differenced out or eliminated due to the nonlinearity of the model. Treating individual heterogeneity as parameters to be estimated leads to the well-known incidental parameter problems where the number of incidental parameters to be estimated increases with the number of individual in the panel leading to inconsistent estimators \parencite{neyman_1948}.

Another important issue concerns whether identification is feasible with short panels and the minimum time periods required for identification. If the panel is long enough, the parameters and unobserved heterogeneity can be estimated consistently. For example, \textcite{Bonhomme_discretizing_2022} propose a method to identify them by discretizing unobserved heterogeneity by grouping individuals using K-means clustering. However, with a short panel, identification becomes difficult due to the limited time periods. For example, in the case of a short panel with two periods, a necessary condition for point identification to be feasible under bounded and strictly exogenous covariates is that the error term follows a logistic distribution \parencite{chamberlain2010binary}.

Several methods have been introduced for identification in nonlinear panel models with strictly exogenous covariates, lagged dependent variables, and unobserved heterogeneity with a fixed time period. \textcite{chamberlain1980} propose a conditional maximum likelihood estimation (CMLE) to address identification in a panel model under a static logit model; the corresponding estimator is consistent and asymptotically normal \parencite{Anderson_1970}. By conditioning on sufficient statistics for the unobserved heterogeneity, the conditional likelihood does not depend on the unobserved heterogeneity.
\textcite{Chamberlain_1985,magnac_2000} provide sufficient statistics under AR(p)-type dynamic logit model. \textcite{Honore_sufficient_2000} further develop the method to accommodate the strictly exogenous variables which are continuous. \textcite{Al-Sadoon21102017} suggest an exponential class specification and related estimation method, CMLE and generalized method of moment, to address identification with lagged variables. \textcite{AGUIRREGABIRIA2021280} extend the approach to structural dynamic logit models with unobserved heterogeneity in the continuation value function accommodating various settings including lagged variables and forward-looking decision-making behavior. \textcite{dano2025binarychoicelogitmodels} study sufficient statistics-based identification strategies for general fixed effects under the binary panel logit model.

Instead of relying on sufficient statistics for the unobserved heterogeneity, which may not exist for some parametric specifications, a generalized method of moments based approach is also introduced. \textcite{bonhomme_functional_2012} proposes a systematic approach to difference out the heterogeneity using the orthogonal projection of the mapping from heterogeneity to the dependent variables. \textcite{KITAZAWA2022350} suggests moment conditions by transforming the dynamic binary panel logit model into linear form. \textcite{dano2023transitionprobabilitiesmomentrestrictions} presents a method to derive moment restrictions in AR(p)-type logit model with unobserved heterogeneity. \textcite{Honore_Moment_functional_2024} extend the approach from \textcite{bonhomme_functional_2012} to a more general setting with lagged variables, providing a numerical method to find explicit analytic expressions for the moment conditions under broader model specifications. Taking advantage of the polynomial structure of the logit-type model with respect to the individual unobserved heterogeneity, \textcite{dobronyi2024identificationdynamicpanellogit} establish the sharp identified set of the parameter, the heterogeneity, and the average treatment effect based on the truncated moment in the dynamic binary panel logit model.

Identified set for the parameter is often derived under weak assumptions. \textcite{honore2006bounds} provide the estimation methods to construct consistent estimates of the identified set. By exploiting the intersection of the inequalities satisfied by the parameters, \textcite{ARISTODEMOU2021253,Khan2023} derive informative bounds and the sharp identified set in the presence of lagged variables and strictly exogenous covariates under minimal assumptions.

Aside from the state dependence, the explanatory variables may not be strictly exogenous, a common assumption in nonlinear panel models, even after controlling for the unobserved heterogeneity as pointed out \textcite{Arellano_Honore_2001}.
As a remedy, several papers address issues of identification and non-identification while relaxing strict exogeneity. For example, \textcite{gao2024identificationnonlineardynamicpanels} provide a general identified set under partial stationarity where there are AR(p)-type of lagged dependent variables and contemporaneously endogenous variables. 

Some papers address issues of non-identification and identification results with predetermined covariates. The covariates may depend on past dependent variables; this dependence is referred to as a feedback process. Contrary to the identification result of \textcite{chamberlain2010binary}, it is not possible to identify the parameters in a short panel with two periods if there exists a lagged dependent variable as a covariate even under logistic error \parencite{Chamberlain_feedback_logit_nonidentification_2023}.  \textcite{bonhomme_identification_2023} provide the conditions for identification under sequential exogeneity and show that even in the simple setup where all covariates are binary, point-identification often fails regardless of the time period. \textcite{bonhomme2025momentrestrictionsnonlinearpanel} characterize feedback and heterogeneity robust moment conditions under sequential exogeneity. 


Identification of the parameters under sequential exogeneity is often possible under the additional assumptions.
Under the assumption that there exists an explanatory variable independent of the unobserved heterogeneity and the errors conditional on the other covariates, \textcite{Honore_Lewbel_2002} establish identification of the other predetermined covariates. \textcite{ARELLANO2003125} present the identification of the parameters through the generalized method of moments when predetermined covariates depend on the conditional expectation of the unobserved heterogeneity. \textcite{PIGINI202283} investigate the identification under the logit model with the assumption that the feedback process belongs to the exponential family via CMLE and pseudo CMLE.



This paper contributes to the literature by exploring identification and non-identification of the parameters in AR(1)-type dynamic logit models with a sequentially exogenous covariate. We show that, even in the presence of a first-order Markovian feedback process of the covariate and individual unobserved heterogeneity, it is possible to construct a conditional likelihood that is free of these components by treating them as nuisance parameters to be eliminated. Under the logit specification, which admits sufficient statistics for the individual heterogeneity, we derive sufficient statistics for the individual unobserved heterogeneity and the feedback process of discrete covariates. By conditioning on the sufficient statistics for both nuisance parameters, the conditional likelihood is free of any individual-specific unobserved parameters. 

Analogous to the identification condition in the standard linear model requiring sufficient variation in the covariates, identification via conditional likelihood often fails because, once conditioning on the sufficient statistics, the conditional likelihood becomes a constant and contains no information about the parameter. This paper relies on this condition to investigate the identification of the parameter via conditional likelihood.

Even beyond identification via conditional likelihood, the identification can fail. This paper establishes a non-identification result under a first order Markovian feedback process by examining the non-existence of the moment function that is a non-constant function of the parameter \parencite{bonhomme_identification_2023,bonhomme2025momentrestrictionsnonlinearpanel}. Failure of point-identification implies that additional restrictions are required to identify $\beta$. This paper suggests the restrictions which are respectively imposed on the feedback process and on the initial condition to identify the parameter.

The rest of the paper is organized as follows. Section \ref{sc:model} presents the AR(1)-type dynamic binary panel logit model, its assumptions, and establishes non-identification results in Subsections \ref{ssc:non_identification1} and \ref{ssc:non_identification2}, which cover non-identification via conditional likelihood and non-identification beyond conditional likelihood. Section \ref{sc:additional_assumption} provides two assumptions to identify the parameters via conditional likelihood in Subsection \ref{ssc:markov_setup2} and \ref{ssc:initial_condition}.  We conclude in Section \ref{sc:conclusion}.




\section{Failure of Point-identification}\label{sc:model}
\subsection{Model}
A sequence of the variable $X$ of individual $i$ from $s$ to $t$ period is denoted by $X_i^{s:t}=\left(X_{is},\hdots,X_{it}\right)$. 
$\left(G(x)\right)_{x\in\mathcal{X}}$ is the collection indexed by the support $\mathcal{X}$ with component $G(x)$ at each $x\in\mathcal{X}$. Cardinality of the set $\mathcal{X}$ is denoted by $|\mathcal{X}|$. The researcher observes panel data $\left(X_{i}^{1:T},Y_{i}^{0:T}\right)$ over $i=1,\hdots,N$ individuals where $Y_{it}\in\{0,1\}$ is a random binary dependent variable and $X_{it}\in\mathcal{X}$ is a random discrete covariate defined on the finite support $\mathcal{X}$. It is assumed that samples are randomly drawn.
\begin{assumption} [Random sampling]\label{as:random_sampling} $(X_i^{1:T},Y_i^{0:T})$ are randomly drawn.
\end{assumption}

Consider the following dynamic panel binary choice model with state dependence of $Y_{it}$ on $Y_{it-1}$, discrete sequentially exogenous $X_{it}$, individual unobserved heterogeneity $\alpha_i$,  and individual time-varying unobserved error term $u_{it}$:
\begin{equation}
Y_{it}=\mathbbm{1}\Big\{\rho Y_{it-1}+\beta X_{it}+\alpha_i\geq u_{it}\Big\},\quad \forall i\in  1,\hdots,N,\quad \forall t=1,\hdots,T.\label{eq:y_specification}
\end{equation}
where $\mathbbm{1}\{\cdot\}$ is the indicator function. Let the parameters of interest be $\theta=\left(\rho,\beta\right)^\prime\in\Theta$ where $\Theta$ denotes the parameter space which is assumed to be a compact subset of $K$ dimensional Eucliden space, $\mathbbm{R}^K$. In the specification \ref{eq:y_specification}, $K=2$.
\begin{assumption}[Compact parameter space]\label{as:compact_parameter}
Parameter space $\Theta$ is a compact subset of the Euclidean space, $\Theta\subset\mathbbm{R}^K$ where $K$ denotes the number of the parameter of interest.
\end{assumption}

In the model specification \ref{eq:y_specification}, the error $u_{it}$ is assumed to follow a logistic distribution given $\left(X_{i}^{1:t},Y_{i}^{0:t-1},\alpha_i\right)$. 
\begin{assumption}[Conditional distribution $u_{it}$]\label{as:stationary} 
$u_{it}$ follows a logistic distribution given $\left(X_{i}^{1:t}, Y_{i}^{0:t-1}, \alpha_i\right)$ 
\begin{equation*}
u_{it}\:|\:X_{i}^{1:t},Y_{i}^{0:t-1},\alpha_i \overset{d}{=} u_{i1}\:|\:X_{i1},Y_{i0},\alpha_i\sim \text{logistic } F,\quad\forall i=  1,\hdots,N,\quad \forall t=2,\hdots,T.
\end{equation*}
\end{assumption} 

Individual unobserved heterogeneity $\alpha_i$ is independent and identically distributed given initial condition $I_i=\iota$. $\alpha_i$ is allowed to be arbitrarily correlated with $\left(X_{i}^{1:T},Y_{i}^{0:T}\right)$. 
\begin{assumption}[Independent and identically distributed $\alpha_i$]\label{as:iida}
$\alpha_i\:|\:I_i=\iota$ is a sequence of independent identically distributed random variables with the probability density function $H_{\iota}\left(\alpha\right)$ and support $\mathcal{A}\subseteq\mathbbm{R}$.
\end{assumption}

Initial condition varies according to the assumptions about $X_{it}$. Under Assumption \ref{as:markov_setup1}, initial conditions are $I_{i}=\left(Y_{i0},X_{i1}\right)$ and the support is $\mathcal{I}=\{0,1\}\times\mathcal{X}$. Under Assumption \ref{as:markov_setup2}, $I_{i}=Y_{i0}$ and the support is $\mathcal{I}=\{0,1\}$. Under Assumption \ref{as:iida_restriction}, $I_{i}=\left(X_{i0},Y_{i0}\right)$ and the support is $\mathcal{I}=\mathcal{X}\times\{0,1\}$.

The objective is to identify $\theta$ and determine the minimum period $T$\footnote{The total number of the time period required for identification is $T+1$ due to the initial condition period; in this paper, we do not count the period $T=0$ for identification.} for identification under Assumptions \ref{as:random_sampling} - \ref{as:iida} while relaxing strict exogeneity on $X_{it}$, which is commonly assumed in the dynamic panel logit model. $X_{it}$ is sequentially exogenous, 
$$
E\left(u_{it}\:|\:X_{i}^{1:t}\right)=0,\quad E\left(u_{it}\:|\:X_{i}^{t+1:T}\right)\neq0,
$$
because of the dependence of $X_{it}$ on the dependent variable in the past. Let the probability  of $X_{it}$ at $x$ given $\left(X_{i}^{t-1},Y_{i}^{t-1}\right)$ be
$$
P\left(X_{it}=x\:|\:X_{i}^{1:t-1},Y_{i}^{0:t-1}\right)=G_i\left(x\:|\:X_{i}^{1:t-1},Y_{i}^{0:t-1}\right).
$$
Note that $X_{it}$ is a discrete random variable. Let this probability of $X_{it}$ be the feedback process of $X_{it}$, which is denoted by $G_i\left(x\:|\:X_{i}^{1:t-1},Y_{i}^{0:t-1}\right)$. The feedback process is assumed to follow the first-order Markovian feedback process in the sense that the process depends only on $\left(X_{it-1},Y_{it-1}\right)$,
$$
G_i\left(x\:|\:X_{i}^{1:t-1},Y_{i}^{0:t-1}\right)=G_i\left(x\:|\:X_{it-1},Y_{it-1}\right),\quad \forall x\in\mathcal{X},\: \forall t\geq2.
$$
This assumption does not require a homogeneous feedback process across individuals. Each individual may have a different feedback process. 

Let $\mathbf{G}_i$ denote a Markov kernel with entries $G_i\left(x_2\:|\:x_1,y\right)$ for all $x_1,x_2\in\mathcal{X}$ and $y\in\{0,1\}$.
\begin{assumption}[Markovian feedback process of $X_{it}$]\label{as:markov_setup1}
Discrete $X_{it}$, defined on a finite support $\mathcal{X}$, follows a first-order Markov process. $\mathbf{G}_i$ is a Markov kernel from $\mathcal{X}\times\{0,1\}$ to the interior of the unit simplex on $\mathcal{X}$, 
$$\mathbf{G}_i:\mathcal{X}\times \{0,1\}\rightarrow \Delta_{+}^\mathcal{X},\quad\mathbf{G}_i\in \mathcal{G}
$$
with entries
$G_i\left(x_2\:|\:x_1,y\right)=P\left(X_{it}=x_2\:|\:X_{it-1}=x_1,Y_{it-1}=y\right)$
for all $y\in\{0,1\}$ and $x_1,x_2\in\mathcal{X}$. $\mathcal{G}$ is the collection of all Markov kernel $\mathbf{G}_i$.
\end{assumption}

By Assumption \ref{as:markov_setup1}, all components of $\mathbf{G}_i$ are  strictly positive. Assumptions \ref{as:compact_parameter} - \ref{as:iida} together with Assumptions about the feedback process of $X_{it}$, \ref{as:markov_setup1} imply that the probability at any sample path $\left(x^{2:T},y^{1:T}\right)$ is strictly positive, 
$$
P\left(X_i^{2:T}=x^{2:T},Y_{i}^{1:T}=y^{1:T}\:|\:Y_{i0},X_{i1},\alpha_i,\mathbf{G}_i\:;\:\theta\right)\in\left(0,1\right),\quad \forall \left(x^{2:T},y^{1:T}\right)\in\mathcal{X}^{T-1}\times \{0,1\}^{T}.
$$
As stated in \textcite{bonhomme_identification_2023}, these assumptions rule out cases of stayers in which $\alpha_i$ and the conditional distribution of $\left(X_{i}^{2:T},Y_{i}^{1:T}\right)$ induce identical values of $X_{it}$ and $Y_{it}$ over time. 

Note that $\mathbf{G}_i$ is an individual-specific feedback process unobserved to the researchers and can be arbitrarily correlated with $\alpha_i$. Therefore, there are two sources of individual heterogeneity, $\alpha_i$ and $\mathbf{G}_i$, which are not observed in the data.
A key identification strategy is to exploit sufficient statistics, treating $\alpha_i$ and $\mathbf{G}_i$ as nuisance parameters to be eliminated. Sufficient statistics are derived using the factorization theorem. The logistic distribution is an example that admits sufficient statistics for $\alpha_i$ and $\mathbf{G}_i$. Under Assumption \ref{as:stationary} - \ref{as:markov_setup1}, the joint probability of $\left(X_i^{2:T},Y_{i}^{1:T}\right)$ at $\left(x^{2:T},y^{1:T}\right)$ given the initial condition $\left(Y_{i0},X_{i1}\right)=\left(y_0,x_1\right)$ in the model specification \ref{eq:y_specification} is
\begin{align}
&P\left(X_i^{2:T}=x^{2:T},Y_{i}^{1:T}=y^{1:T}\:|\:Y_{i0}=y_0,X_{i1}=x_1,\alpha_i,\mathbf{G}_i\:;\:\theta\right)\nonumber\\
&=\prod_{t=1}^T P\left(Y_{it}=y_t\:|\:y_{t-1},x_{t},\alpha_i\:;\:\theta \right)\prod_{t=2}^T G_i\left(x_t\:|\:x_{t-1},y_{t-1}\right)\nonumber\\
&=\prod_{t=1}^T F\left(\rho y_{t-1}+\beta x_{t}+\alpha_i\right)^{y_{t}} \left(1- F\left(\rho y_{t-1}+\beta x_{t}+\alpha_i\right)\right)^{1-{y_{t}}}\prod_{t=2}^T G_i\left(x_t\:|\:x_{t-1},y_{t-1}\right)\nonumber\\
&=\underbrace{\exp{\left(\rho \sum_{t=1}^T y_{t-1}y_{t}+\beta \sum_{t=1}^T x_{t}y_{t}\right)}}_{f_{\theta}\left(x^{1:T},y^{0:T}\right)}\underbrace{\frac{\exp\left(\alpha_i\sum_{t=1}^T y_{t}\right)}{\prod_{t=1}^T\Big\{1+\exp{\left(\rho y_{t-1}+\beta x_{t}+\alpha_i\right)}\Big\}}}_{f_{\alpha_i}\left(x^{1:T},y^{0:T}\right)}\underbrace{\prod_{t=2}^T G_i\left(x_t\:|\:x_{t-1},y_{t-1}\right)}_{f_{\mathbf{G}_i}\left(x^{1:T},y^{1:T-1}\right)}\label{eq:logit_setup1}
\end{align}
where the first equality holds due to the model specification from equation (\ref{eq:y_specification}) and Assumption \ref{as:stationary}. The second equality follows from the fact that $Y_{it}$ is binary, and the third equality holds because the conditional distribution $F$ is the logistic distribution function. 

The joint probability at $\left(x^{1:T},y^{0:T}\right)$ can be factorized into functions $f_{\theta}$, $f_{\alpha_i}$, and $f_{\mathbf{G}_i}$. $f_{\alpha_i}$ and $f_{\mathbf{G}_i}$ are functions of nuisance parameters $\alpha_i$ and $\mathbf{G}_i$, respectively, while $f_{\theta}$ is free of nuisance parameters. By the discrete support of $X_{it}$ and $Y_{it-1}$,  the denominator in $f_{\alpha_i}$, 
$\prod_{t=1}^T \Big\{1+\exp{\left(\rho y_{t-1}+\beta x_{t}+\alpha_i\right)}\Big\}$ 
can be expressed in terms of the counts of possible values taken by $\Big\{1+\exp{\left(\rho y_{t-1}+\beta x_{t}+\alpha_i\right)}\Big\}$:
\begin{align}
f_{\alpha_i}\left(x^{1:T},y^{0:T}\right)&=\frac{\exp\left(\alpha_i\sum_{t=1}^T y_{t}\right)}{\prod_{t=1}^T\Big\{1+\exp{\left(\rho y_{t-1}+\beta x_{t}+\alpha_i\right)}\Big\}}\nonumber\\
&=\frac{\exp\left(\alpha_i\sum_{t=1}^T y_{t}\right)}{\prod_{y\in\{0,1\}} \prod_{x\in\mathcal{X}}\Big\{1+\exp{\left(\rho y+\beta x+\alpha_i\right)}\Big\}^{\sum_{t=1}^{T} \mathbbm{1}\{y_{t-1}=y,x_{t}=x\}}}.\label{eq:f_2}
\end{align}
$f_{\alpha_i}$ can be represented as a function of the statistics related to $\alpha_i$. Let the vector of the statistics which indicates the number of transitions of $X_{it}$ given $Y_{it-1}$, $\sum_{t=1}^{T} \mathbbm{1}\{Y_{it-1}=y,X_{it}=x\}$, for all $x\in\mathcal{X},y\in\{0,1\}$ be $\mathbf{S}_1^{\alpha}$, 
$$
\mathbf{S}_1^{\alpha}\left(X_i^{1:T},Y_{i}^{0:T-1}\right)=\left(\sum_{t=1}^{T} \mathbbm{1}\{Y_{it-1}=y,X_{it}=x\}\right)_{x\in\mathcal{X},y\in\{0,1\}}.
$$
The denominator of $f_{\alpha_i}$ is a function of $\left(X_{i}^{1:T},Y_{i}^{0:T-1}\right)$ via $\mathbf{S}_1^{\alpha}$. Based on the binary $Y_{it}$ and discrete $X_{it}$, if the number of all possible pairs $\left(Y_{it-1},X_{it}\right)$ in $(X_{i}^{1:T},Y_{i}^{0:T-1})$ is known given initial condition $\left(Y_{i0},X_{i1}\right)=\left(y_0,x_1\right)$, the denominator of $f_{\alpha_i}$ is known. Similarly, if $\sum_{t=1}^T Y_{it}$ is known, the numerator in $f_{\alpha_i}$ is known, which shows that the numerator of $f_{\alpha_i}$ depends on $\alpha_i$ through statistics $\sum_{t=1}^T Y_{it}$. Therefore, $f_{\alpha_i}$ depends on $\alpha_i$ through the statistics $\mathbf{S}^\alpha$,
$$
\mathbf{S}^{\alpha}\left(X_i^{1:T},Y_{i}^{0:T}\right)=\left(\sum_{t=1}^T Y_{it},\mathbf{S}_1^{\alpha}\right).
$$
$f_{\alpha_i}\left(x^{1:T},y^{0:T}\right)$ can be expressed as
$$
f_{\alpha_i}\left(x^{1:T},y^{0:T}\right)=f_{\alpha_i}\left(s^\alpha\right)
$$
where $s^\alpha$ denotes the value of $\mathbf{S}^\alpha$ at $\left(x^{1:T},y^{0:T}\right)$.


As an analogy of the representation of $f_{\alpha_i}$ using $\mathbf{S}^{\alpha}$, $f_{\mathbf{G}_i}$ can be represented in terms of the counts of possible cases in $G_i\left(x_t\:|\:x_{t-1},y_{t-1}\right)$:
\begin{align}
f_{\mathbf{G}_i}\left(x^{1:T},y^{1:T-1}\right)&=\prod_{t=2}^TG_i\left(x_t\:|\:x_{t-1},y_{t-1}\right)\nonumber\\
&=\prod_{y\in\{0,1\}}\prod_{x_1,x_2\in\mathcal{X}}G_i\left(x_2\:|\:x_1,y\right)^{\sum_{t=1}^{T-1} \mathbbm{1}\{x_{t}=x_1,y_{t}=y,x_{t+1}=x_2\}}.\label{eq:setup1_f3}
\end{align}
Note that this representation holds regardless of the conditional distribution of $u_{it}$. $f_{\mathbf{G}_i}$ depends on $\mathbf{G}_i$ through a vector of statistics $\mathbf{S}^{\mathbf{G}}$ whose components are the number of all possible triples $\left(X_{it},Y_{it},X_{it+1}\right)$,
$$
\mathbf{S}^{\mathbf{G}}\left(X_i^{1:T},Y_{i}^{1:T-1}\right)=\left(\sum_{t=1}^{T-1} \mathbbm{1}\{X_{it}=x_1,Y_{it}=y,X_{it+1}=x_2\}\right)_{x_1,x_2\in\mathcal{X},y\in\{0,1\}}.
$$
Consequently,
$$
f_{\mathbf{G}_i}\left(x^{1:T},y^{1:T-1}\right)=f_{\mathbf{G}_i}\left(s^\mathbf{G}\right).
$$
where $s^\mathbf{G}$ denotes the value of $\mathbf{S}^\mathbf{G}$ at $\left(x^{1:T},y^{1:T-1}\right)$. First-order Markovian feedback Assumption \ref{as:markov_setup1} allows the proposed approaches to be applied.

Therefore, the joint probability in (\ref{eq:logit_setup1}) can be factorized into functions of the statistics $\mathbf{S}^\alpha$ and $\mathbf{S}^{\mathbf{G}}$, and a function that is free of $\alpha_i$ and $\mathbf{G}_i$,
\begin{equation}
f_{\theta}\left(x^{1:T},y^{0:T}\right)f_{\alpha_i}\left(x^{1:T},y^{0:T}\right)f_{\mathbf{G}_i}\left(x^{1:T},y^{1:T-1}\right)=f_{\theta}\left(x^{1:T},y^{0:T}\right)f_{\alpha_i}\left(s^\alpha\right)f_{\mathbf{G}_i}\left(s^\mathbf{G}\right).\label{eq:setup1_factorization}
\end{equation}
$f_\theta$ and $f_{\alpha_i}$ are strictly positive functions, and $f_{\mathbf{G}_i}$ is strictly positive according to Assumption \ref{as:markov_setup1}. Thus, by the factorization theorem,
\begin{align}
\mathbf{S}\left(X_i^{1:T},Y_{i}^{0:T}\right)&=\left(\mathbf{S}^\alpha\left(X_i^{1:T},Y_{i}^{0:T}\right),\mathbf{S}^\mathbf{G}\left(X_i^{1:T},Y_{i}^{1:T-1}\right)\right)\nonumber\\
&=\left(\sum_{t=1}^TY_{it},\left(\sum_{t=1}^{T-1} \mathbbm{1}\{X_{it}=x_1,Y_{it}=y,X_{it+1}=x_2\}\right)_{x_1,x_2\in\mathcal{X},y\in\{0,1\}}\right)\label{eq:suff_model1}
\end{align}
are sufficient statistics for the nuisance parameters $\alpha_i$ and $\mathbf{G}_i$. Observe that the statistic $\mathbf{S}^\mathbf{G}$ imposes more constraints in $\left(X_{i}^{1:T},Y_{i}^{1:T-1}\right)$ than $\mathbf{S}^{\alpha}$ since $\mathbf{S}^{\mathbf{G}}$ counts the possible triples $\left(X_{it},Y_{it},X_{it+1}\right)$, while $\mathbf{S}^{\alpha}$ counts the possible pairs $\left(Y_{it-1},X_{it}\right)$. If $\mathbf{S}^\mathbf{G}$ is known, $\mathbf{S}^{\alpha}$ is determined. This restriction on possible $\left(X_i^{2:T},Y_{i}^{1:T}\right)$ is the source of non-identification of $\beta$ through the conditional likelihood under Assumption \ref{as:markov_setup1}.

\begin{remark}
Analogous to the identification strategies in \textcite{ARELLANO2003125,bonhomme_identification_2023} where all possible values of the feedback process of $X_{it}$ in each period are treated as parameters, we treat the feedback process associated with each possible triple $\left(y_{t-1},x_{t-1},x_{t}\right)\in \{0,1\}\times\mathcal{X}^2$ as a nuisance parameter. Since the feedback process follows a first-order Markov chain, there are $2|\mathcal{X}|^2$ parameters related to the feedback process under Assumption \ref{as:markov_setup1}. Under Assumption \ref{as:markov_setup2}, this number reduces to $2|\mathcal{X}|$. The distinguishing feature compared to \textcite{ARELLANO2003125,bonhomme_identification_2023} is that, under first-order Markov feedback, the number is invariant to the time period $T$, which enables the construction of sufficient statistics for the feedback process.
\end{remark}

\subsection{Failure of Point-identification via Conditional Likelihood}\label{ssc:non_identification1}
Let $\mathcal{S}$ denote the support of $\mathbf{S}$. For each $s\in\mathcal{S}$, let $\mathcal{D}_s$ denote the set of possible values of $(x^{2:T},y^{1:T})$ such that $\mathbf{S}\left(x^{1:T},y^{0:T}\right)=s$:
\begin{equation}
\mathcal{D}_s=\Big\{\left(x^{2:T},y^{1:T}\right):\mathbf{S}\left(x^{1:T},y^{0:T}\right)=s\Big\}.\label{eq:history_set_setup1}
\end{equation}
From equation (\ref{eq:logit_setup1}) and (\ref{eq:setup1_factorization}), the probability of $\left(X_i^{2:T},Y_{i}^{1:T}\right)$ at $\left(x^{2:T},y^{1:T}\right)$ conditional on $\mathbf{S}=s$ and the initial condition $\left(Y_{i0},X_{i1}\right)=\left(y_0,x_1\right)$ is
\begin{align}
&P\left(X_i^{2:T}=x^{2:T},Y_{i}^{1:T}=y^{1:T}\:|\:\mathbf{S}=s,Y_{i0}=y_0,X_{i1}=x_1\:;\:\theta\right)\nonumber\\
&=P\left(X_i^{2:T}=x^{2:T},Y_{i}^{1:T}=y^{1:T}\:|\:\left(X_i^{2:T},Y_i^{1:T}\right)\in \mathcal{D}_s,Y_{i0}=y_0,X_{i1}=x_1\:;\:\theta\right)\nonumber\\
&=\frac{\exp{\left(\rho \sum_{t=1}^T y_{t-1} y_{t}+\beta\sum_{t=1}^T x_{t}y_{t}\right)}}{\sum_{\left(\tilde{x}^{2:T},\tilde{y}^{1:T}\right)\in \mathcal{D}_s}\exp{\left(\rho \sum_{t=1}^T \tilde{y}_{t-1} \tilde{y}_{t}+\beta\sum_{t=1}^T \tilde{x}_{t} \tilde{y}_{t}\right)}},\quad \forall s \in\mathcal{S}.\label{eq:conditional_likelihood}
\end{align}
where $\tilde{x}_{t},\tilde{y}_{t}$ denotes $x_{t},y_{t}$ in $(\tilde{x}^{2:T},\tilde{y}^{1:T})$ and $\left(\tilde{y}_{0},\tilde{x}_{1}\right)=\left(y_0,x_1\right)$. Note that the probability (\ref{eq:conditional_likelihood}) does not depend on $\alpha_i$ and $\mathbf{G}_i$.

Identification of parameters of interest via conditional likelihood relies on the conditional probability (\ref{eq:conditional_likelihood}). Let $\theta_0$ denote the true value of the parameters.
The identification is equivalent to verifying whether there exists $\mathbf{S}=s$ such that
\begin{multline}
P\left(X_i^{2:T}=x^{2:T},Y_{i}^{1:T}=y^{1:T}\:|\:\left(X_i^{2:T},Y_i^{1:T}\right)\in \mathcal{D}_s,Y_{i0}=y_0,X_{i1}=x_1\:;\:\theta_0\right)\\
=P\left(X_i^{2:T}=x^{2:T},Y_{i}^{1:T}=y^{1:T}\:|\:\left(X_i^{2:T},Y_i^{1:T}\right)\in \mathcal{D}_s,Y_{i0}=y_0,X_{i1}=x_1\:;\:\theta_1\right)\label{eq:identification_condition}
\end{multline}
is equivalent to $\theta_0=\theta_1$. The corresponding identification condition from equation (\ref{eq:identification_condition}) is
\begin{multline*}
\sum_{(\tilde{x}^{2:T},\tilde{y}^{1:T})\in \mathcal{D}_s}\exp{\left(\rho_1 \sum_{t=1}^T \tilde{y}_{t-1} \tilde{y}_{t}+\beta_1\sum_{t=1}^T \tilde{x}_{t} \tilde{y}_{t}+\rho_0 \sum_{t=1}^T y_{t-1} y_{t}+\beta_0\sum_{t=1}^T x_{t}y_{t}\right)}\\
=\sum_{(\tilde{x}^{2:T},\tilde{y}^{1:T})\in \mathcal{D}_s}\exp{\left(\rho_0 \sum_{t=1}^T \tilde{y}_{t-1} \tilde{y}_{t}+\beta_0\sum_{t=1}^T \tilde{x}_{t} \tilde{y}_{t}+\rho_1 \sum_{t=1}^T y_{t-1} y_{t}+\beta_1\sum_{t=1}^T x_{t}y_{t}\right)}.
\end{multline*}
To hold equality, for all possible $(\tilde{x}^{1:T},\tilde{y}^{0:T})$ conditional on $\mathbf{S}=s$,
\begin{equation}
\left(\rho_0-\rho_1\right)\left(\sum_{t=1}^T \tilde{y}_{t-1} \tilde{y}_{t}-\sum_{t=1}^T y_{t-1} y_{t}\right)+\left(\beta_0-\beta_1\right)\left(\sum_{t=1}^T \tilde{x}_{t} \tilde{y}_{t}-\sum_{t=1}^T x_{t}y_{t}\right)=0, \label{eq:identification_condition_logit}
\end{equation}
holds. For all $s$ in $\mathcal{S}$, if there does not exist $(\tilde{x}^{1:T},\tilde{y}^{0:T})$ conditional on $\mathbf{S}=s$ such that 
\begin{align}
&\text{(for $\rho$)}\quad\sum_{t=1}^T \tilde{y}_{t-1} \tilde{y}_{t}\neq\sum_{t=1}^T y_{t-1} y_{t},
&\text{(for $\beta$)}\quad\sum_{t=1}^T \tilde{x}_{t} \tilde{y}_{t}\neq\sum_{t=1}^T x_{t}y_{t}\label{eq:identifying_stat}
\end{align}
with positive probability, identification of the corresponding parameter via conditional likelihood fails because even if $\rho_0\neq\rho_1$ or $\beta_0\neq\beta_1$, equation (\ref{eq:identification_condition_logit}) can still hold. If every possible $(\tilde{x}^{2:T},\tilde{y}^{1:T})$ conditioning on $\mathbf{S}=s$ has identical value of both $\sum_{t=1}^T \tilde{y}_{t-1} \tilde{y}_{t}$ and $\sum_{t=1}^T \tilde{x}_{t} \tilde{y}_{t}$ or if $\mathcal{D}_s$ is a singleton, equation (\ref{eq:identification_condition}) holds. In this case, the distributions of $\sum_{t=1}^T Y_{it-1} Y_{it}$ and $\sum_{t=1}^T X_{it} Y_{it}$ conditional on $\mathbf{S}=s$ are degenerate. The probability (\ref{eq:conditional_likelihood}) is a constant and holds for any parameter value.
Therefore, identification of the parameters fails. Consequently, a key to the identification of parameters through conditional likelihood is to verify whether there exists $s$ such that condition (\ref{eq:identifying_stat}) holds once conditional on $\mathbf{S}=s$. 

\begin{remark}
Suggested in \textcite{DHAULTFOEUILLE201677,dano2025binarychoicelogitmodels}, identification via conditional likelihood is equivalent to finding permutations of $\left(X_{i}^{2:T},Y_{i}^{1:T}\right)$ under restrictions $\mathbf{S}$ given $\left(Y_{i0},X_{i1}\right)$ and verifying whether these permutations generate different values of $\sum_{t=1}^T Y_{it-1} Y_{it}$ and $\sum_{t=1}^T X_{it} Y_{it}$. If all permutations produce identical values of these statistics, there is no variation and identification via conditional likelihood fails.
\end{remark}

\begin{remark}
The condition (\ref{eq:identifying_stat}) is in line with the identification of CMLE under the logit specification with strictly exogenous $X_{it}$ and fixed effect described in \textcite{davezies2024identificationestimationaveragecausal,dano2025binarychoicelogitmodels}. As shown in Proposition \ref{pp:setup1} in \textcite{davezies2024identificationestimationaveragecausal}, if there exist enough within-unit time variations in $X$, the expected second derivative of conditional likelihood $l_c\left(y|x,\beta\right)$,
$$
E\Bigg[\frac{\partial^2}{\partial\beta\partial \beta^\prime}l_c\left(y|x,\beta\right)\Bigg]=E\Bigg[\frac{\partial^2}{\partial\beta\partial \beta^\prime} \left( \sum_{t=1}^Ty_tx_t^\prime \beta-\ln \sum_{\left(d_1,\hdots,d_T\right)\in\{0,1\}^{T}:\sum_{t=1}^{T}d_t=k}\exp\left(\sum_{t=1}^T d_tx_t^\prime\beta\right)\right)\Bigg]
$$
is negative definite. Therefore, the population CMLE objective has a unique maximizer. 
Under strictly exogenous $X_{it}$ with lagged outcome $Y_{it-1}$, and generalized fixed effect, Theorems 1 and 2 in \textcite{dano2025binarychoicelogitmodels} also provide analogous conditions for identification that require sufficient variations in $\sum_{t=1}^TX_{it}Y_{it}$ and $\sum_{t=1}^TY_{it-1}Y_{it}$ which are statistics related to the parameters of interest. As an extension of the case with sequentially exogenous $X_{it}$, suppose that for some realization of $s_i$ of the sufficient statistics $\mathbf{S}_i=\mathbf{S}\left(X_{i}^{1:T},Y_{i}^{0:T}\right)$, there is variation in $\sum_{t=1}^TX_{it}Y_{it}$ and $\sum_{t=1}^TY_{it-1}Y_{it}$ across the sample paths consistent with $\mathbf{S}_i=s_i$. Then, the CMLE sample objective function,
\begin{align*}
&\hat{Q}_N\left(\theta\right)=\frac{1}{N}\sum_{i=1}^N\ln P\left(X_i^{2:T}=x_i^{2:T},Y_{i}^{1:T}=y_i^{1:T}\:|\:\mathbf{S}_i=s_i,Y_{i0},X_{i1}\:;\:\theta\right),
\end{align*}
has a unique maximizer; the population objective function,
\begin{align*}
Q_N\left(\theta\right)=\frac{1}{N}\sum_{i=1}^NE_{\theta,\mathbf{G}_i}\Big[\ln P\left(X_i^{2:T}=x_i^{2:T},Y_{i}^{1:T}=y_i^{1:T}\:|\:\mathbf{S}_i,Y_{i0},X_{i1}\:;\:\theta\right)\Big],
\end{align*}
has a unique maximizer.\footnote{Following the notation in \textcite{Anderson_1970}, let the probability of the sufficient statistics be $P(\mathbf{S}_i=s_i\:|\:I_i,\alpha_i,\mathbf{G}_i\:;\:\theta)=g_\theta(s_i\:|\:I_i,\alpha_i,\mathbf{G}_i)$. $E_{\theta}\left(f\:|\:\:\mathbf{S}_i=s_i,I_i\right)$ is an expectation of $f$ with respect to the probability (\ref{eq:conditional_likelihood}). $E_{\theta,\alpha_i,\mathbf{G}_i}\left(f\:|\:I_i\right)$ is the expectation of $f$ over $g_\theta(s_i\:|\:I_i,\alpha_i,\mathbf{G}_i)$. $E_{\theta,\mathbf{G}_i}\left(f\:|\:I_i\right)$ is an integrated expectation of $E_{\theta,\alpha_i,\mathbf{G}_i}\left(f\right)$ over $\alpha_i\:|\:I_i$, $E_{\theta,\mathbf{G}_i}\left(f\:|\:I_i\right)=\int_{\alpha}E_{\theta,\alpha,\mathbf{G}_i}\left(f\right)H_{I_i}\left(\alpha\right)d\alpha$.} In Appendix \ref{ap:many_proof}, we show that the maximizer is unique for each objective function.
\end{remark}

Proposition \ref{pp:setup1} establishes the identification result via conditional likelihood in the model exploiting the condition (\ref{eq:identifying_stat}).
\begin{restatable}{proposition}{PropnonidentificationCMLE}
\label{pp:setup1}
Consider following dynamic logit model specification under Assumptions \ref{as:compact_parameter} - \ref{as:markov_setup1}:
\begin{multline*}
P\left(X_i^{2:T}=x^{2:T},Y_{i}^{1:T}=y^{1:T}\:|\:Y_{i0}=y_0,X_{i1}=x_1,\alpha_i,\mathbf{G}_i\:;\:\theta\right)
\\=\exp{\left(\rho \sum_{t=1}^T y_{t-1}y_{t}+\beta \sum_{t=1}^T x_{t}y_{t}\right)}
\frac{\exp\left(\alpha_i\sum_{t=1}^T y_{t}\right)}{\prod_{t=1}^T\Big\{1+\exp{\left(\rho y_{t-1}+\beta x_{t}+\alpha_i\right)}\Big\}} \prod_{t=2}^T G_i\left(x_t\:|\:x_{t-1},y_{t-1}\right).
\end{multline*}
For any $T$, identification of $\beta$ via conditional likelihood fails. For $T\geq3$, $\rho$ is identified via conditional likelihood.
\end{restatable}
For the proof of  Proposition \ref{pp:setup1}, see Appendix \ref{ap:proof}. As an example, consider following two sample paths of $\left(X_i^{2:T},Y_{i}^{1:T}\right)$ given $\left(Y_{i0},X_{i1}\right)=\left(0,1\right)$ for the identification of $\rho$ under $T=3$:
$$
\begin{array}{ccccc}
&Y_{i0}&(X_{i1},Y_{i1}) &(X_{i2},Y_{i2})&(X_{i3},Y_{i3}) \\\hline\hline
A:&0 & (1,0) &(1,1)&(1,1) \\
B:&0 & (1,1) &(1,0)&(1,1)
\end{array}
$$
where $A$ and $B$ have identical values of the sufficient statistics $\mathbf{S}$. They have an identical number of occurrences of the feedback process $(X_{it},Y_{it},X_{it+1})$ over periods: $(1,0,1)$, $(1,1,1)$.
In addition, $\sum_{t=1}^T Y^A_{it}=\sum_{t=1}^T Y^B_{it}=2.$
In terms of $\sum Y_{it-1} Y_{it}$, $A$ and $B$ have different values,
$\sum_{t=1}^T Y_{it-1}^A Y_{it}^A=1$, $\sum_{t=1}^T Y_{it-1}^B Y_{it}^B=0$. Since there exists $\mathbf{S}=s$ and identification condition (\ref{eq:identifying_stat}) is satisfied, $\rho$ is identified.

Once conditional on the sufficient statistics \ref{eq:suff_model1}, there do not exist sufficient variations in the value of $\sum_{t=1}^TX_{it}Y_{it}$. There might exist multiple paths conditional on $\mathbf{S}=s$. However, all paths have an identical value of $\sum_{t=1}^TX_{it}Y_{it}$, violating the identification condition (\ref{eq:identifying_stat}). Conditioning on the sufficient statistics eliminates all variation in $\sum_{t=1}^TX_{it}Y_{it}$.

Non-identification result of $\beta$ through conditional likelihood does not imply that $\beta$ is not identified in general. CMLE only exploits the information about the parameters from the conditional likelihood (\ref{eq:conditional_likelihood}), not from the joint probability (\ref{eq:logit_setup1}). Moreover, there exist distributions $F$ that do not admit sufficient statistics for $\alpha_i$; hence, identification of the parameter conditioning on the sufficient statistics for $\alpha_i$ and $\mathbf{G}_i$ is not always plausible method for identification. Therefore, identification via conditional likelihood does not provide a generalized condition for identification. 
Section \ref{ssc:non_identification2} provides failure of point-identification under the Markovian process in panel logit models by examining non-existence of feedback and heterogeneity robust moment conditions described in \textcite{bonhomme_identification_2023,bonhomme2025momentrestrictionsnonlinearpanel}.


\subsection{Failure of Point-identification beyond Conditional Likelihood}\label{ssc:non_identification2}
Analogous to the model specification in \textcite{bonhomme_identification_2023}, to illustrate the non-identification of $\beta$ via observed probabilities, consider the following simplified model with a binary sequentially exogenous $X_{it}$, individual unobserved heterogeneity $\alpha_i$, and individual time-varying unobserved error term $u_{it}$:
\begin{equation}
Y_{it}=\mathbbm{1}\Big\{\beta X_{it}+\alpha_i\geq u_{it}\Big\},\quad \forall i\in  1,\hdots,N,\quad \forall t=1,\hdots,T.\label{eq:y_specification_simple}
\end{equation}
For simplicity, suppose the support of $\alpha_i$ is finite and contains more than $2^{2T-1}-1$ points\footnote{If the number of points in the support is less than $2^{2T-1}-1$, each point in the support would be pinned down as a solution to a nonlinear system of equations. In this case, the rank of the matrix $\mathbf{P}$ in the equation (\ref{eq:prob_matrix}) is 
$$
\rank\left(\mathbf{P}\right)\leq \min\left(|\mathcal{A}|,2^{2T-1}-1\right)=|\mathcal{A}|.
$$
To avoid this, it is assumed that the support $\mathcal{A}$ contains more than $2^{2T-1}-1$.},
$$
\mathcal{A}=\{\alpha_1,\hdots,\alpha_K\},\quad|\mathcal{A}|>2^{2T-1}-1.
$$

Identical to the feedback process specification in \textcite{bonhomme_identification_2023}, for simplicity, assume that heterogeneity in $\mathbf{G}_i$ enters only through $\alpha_i$: 
$$
G_{i}\left(1\:|\:x,y\right)=G\left(1\:|\:x,y\:;\:\alpha_i\right)=G_{xy}\left(\alpha_i\right), \quad \forall x,y\in\{0,1\}
$$
where $G_{xy}\left(\alpha_i\right)$ is shorthand for the feedback process evaluated at $\alpha_i$. Denote the value of the sufficient statistics $\mathbf{S}$ from (\ref{eq:suff_model1}) of the probability of the path $j$ by
$$
\left(n_y^j,\left(n_{x_1,x_2,y}^j\right)_{x_1,x_2,y\in\{0,1\}}\right).
$$
and the exponential of $\alpha_i$ by $A_i$,
$$
A_i=\exp\left(\alpha_i\right).
$$
Using the notations, the integrated likelihood of $\left(X_i^{2:T},Y_{i}^{1:T}\right)$ at $j$ sample path $\left(x^{j,2:T},y^{j,1:T}\right)$ over $\alpha_i$ is
\begin{align*}
&P\left(X_i^{2:T}=x^{j,2:T},Y_{i}^{1:T}=y^{j,1:T}\:|\:X_{i1}=x_1,\mathbf{G}\:;\:\beta\right)
=\sum_{\alpha_k\in\mathcal{A}}p_j\left(\alpha_k\right)H_{x_1}\left(\alpha_k\right)
\\
&=\sum_{\alpha_k\in\mathcal{A}}\exp{\left(\beta \sum_{t=1}^T x_{t}^jy_{t}^j\right)}
A_k^{n_y^j}\Big\{1+\exp{\left(\beta\right)A_k\Big\}^{-\sum_{x,y\in\{0,1\}}n_{x,1,y}^{j}}\Big\{1+A_k}\Big\}^{-\sum_{x,y\in\{0,1\}}n_{x,0,y}^{j}} \\
&\times\Big\{1+\exp\left(\beta x_1\right)A_k\Big\}^{-1}\prod_{x,y\in\{0,1\}}
G_{xy}^{n_{x,1,y}^{j}}\left(\alpha_k\right)\left(1-G_{xy}\left(\alpha_k\right)\right)^{n_{x,0,y}^{j}}H_{x_1}\left(\alpha_k\right).
\end{align*}
where $p_j\left(\alpha_k\right)$ is the shorthand notation of the conditional probability of $j$ sample paths given $\alpha_k$ and $\mathbf{G}=\left(G_{xy}\left(\alpha_k\right)\right)_{x,y\in\{0,1\},\alpha_k\in\mathcal{A}}$. There are $2^{2T-1}$ sample paths. Let $J+1=2^{2T-1}$ and $J$ paths need to be analyzed due to the fact that the sum of the probabilities of all sample paths is one.

Parameters in the model are structural parameter $\beta$, probability mass function of $\alpha_i$ at each point in the support $\mathcal{A}$, and feedback process of $X_{it}$,
$\left(\beta,\mathbf{H}_{x_1},\mathbf{G}\right)$
where $\mathbf{H}_{x_1}=\left(H_{x_1}\left(\alpha_k\right)\right)_{\alpha_k\in\mathcal{A}}$. Let the vector of the integrated probabilities of all sample paths given $X_{i1}=x_1$ be $\psi_{x_1}\left(\beta,\mathbf{H}_{x_1},\mathbf{G}\right)$,
$$
\psi_{x_1}\left(\beta,\mathbf{H}_{x_1},\mathbf{G}\right)
=\begin{pmatrix}
\sum_{\alpha_k\in\mathcal{A}}p_1\left(\alpha_k\right)H_{x_1}\left(\alpha_k\right)\\
\vdots
\\
\sum_{\alpha_k\in\mathcal{A}}p_J\left(\alpha_k\right)H_{x_1}\left(\alpha_k\right)
\end{pmatrix}
$$ 
and let the corresponding Jacobian matrix be 
$$
\nabla \psi_{x_1}\left(\beta,\mathbf{H}_{x_1},\mathbf{G}\right)=
\begin{pmatrix}
    \nabla_{\beta} \psi_{x_1}\left(\beta,\mathbf{H}_{x_1},\mathbf{G}\right) & \nabla_{\mathbf{H}_{x_1}} \psi_{x_1}\left(\beta,\mathbf{H}_{x_1},\mathbf{G}\right) & \nabla_{\mathbf{G}} \psi_{x_1}\left(\beta,\mathbf{H}_{x_1},\mathbf{G}\right)
\end{pmatrix}.
$$

Identification requires the rank of the Jacobian matrix $\nabla \psi_{x_1}\left(\beta,\mathbf{H}_{x_1},\mathbf{G}\right)$ evaluated at the true parameter value, which is not attainable because the true value of the parameter is unknown. Regularity at the true parameter values ensures that the rank is locally constant in a neighborhood of the truth, so it is possible to compute the rank near the truth without knowing the exact true parameter.
\begin{assumption}\label{as:regular_point}
$\left(\beta,\mathbf{H}_{x_1},\mathbf{G}\right)$ is a regular point of the Jacobian matrix $\nabla \psi_{x_1}\left(\beta,\mathbf{H}_{x_1},\mathbf{G}\right)$. That is, the rank of $\nabla \psi_{x_1}\left(\beta,\mathbf{H}_{x_1},\mathbf{G}\right)$ is constant for all points in an open neighborhood of $\left(\beta,\mathbf{H}_{x_1},\mathbf{G}\right)$.
\end{assumption} 
As pointed out in \textcite{bonhomme_identification_2023,bekker2001identification}, Assumption \ref{as:regular_point} is satisfied almost everywhere in the logit specification since the cumulative distribution of the logistic $F$ is an analytic function.

The parameter $\beta$ under the parametric models is identified if 
$$
\rank\left(\nabla_{-\beta} \psi_{x_1}\left(\beta,\mathbf{H}_{x_1},\mathbf{G}\right)\right)<\rank\left(\nabla \psi_{x_1}\left(\beta,\mathbf{H}_{x_1},\mathbf{G}\right)\right)
$$
where $\nabla_{-\beta} \psi_{x_1}\left(\beta,\mathbf{H}_{x_1},\mathbf{G}\right)$ is the Jacobian matrix except $\beta$ column. Equivalently, if one can show 
$$
\rank\left(\nabla_{-\beta} \psi_{x_1}\left(\beta,\mathbf{H}_{x_1},\mathbf{G}\right)\right)=\rank\left(\nabla \psi_{x_1}\left(\beta,\mathbf{H}_{x_1},\mathbf{G}\right)\right),
$$
the parameter $\beta$ is not identified \parencite{rothenberg1971identification,bekker2001identification}. Exploiting the conditions, \textcite{bonhomme_identification_2023} investigates the necessary condition for identification of $\beta$ in dynamic panel logit models that if $\beta$ is identified, there exists a nonzero projection $m^{proj}$ of $\nabla_{\beta} \psi_{x_1}$ onto the orthogonal complement of the space spanned by $\begin{pmatrix}
\nabla_{\mathbf{H}_{x_1}} \psi_{x_1} & \nabla_{\mathbf{G}} \psi_{x_1}
\end{pmatrix}$ such that
$$
m^{proj}= \underbrace{\left(I-\begin{pmatrix}
\nabla_{\mathbf{H}_{x_1}} \psi_{x_1} & \nabla_{\mathbf{G}} \psi_{x_1}
\end{pmatrix}\begin{pmatrix}
\nabla_{\mathbf{H}_{x_1}} \psi_{x_1} & \nabla_{\mathbf{G}} \psi_{x_1}
\end{pmatrix}^\dagger\right)}_{\mathbf{M}}\nabla_{\beta}\psi_{x_1}
$$
where $\dagger$ denotes Moore-Penrose inverse matrix. By the properties of the projection which is nonzero, it follows that
\begin{equation}
\nabla_{\beta} \psi_{x_1}^\prime m^{proj}\neq0,\quad \nabla_{\mathbf{H}_{x_1}} \psi_{x_1}^\prime m^{proj}=0,\quad \nabla_{\mathbf{G}} \psi_{x_1}^\prime m^{proj}=0.\label{eq:identification_moment_require}
\end{equation}
Notice that since $m^{proj}$ is nonzero, $m^{proj}$ satisfies $\nabla_\beta \psi_{x_1}^\prime m^{proj}\neq0$.\footnote{Suppose $\nabla_\beta \psi_{x_1}^\prime m^{proj}=0$ holds. Since $m^{proj}=\mathbf{M}\nabla_{\beta}\psi_{x_1}$, $\nabla_{\beta} \psi_{x_1}^\prime m^{proj}=\nabla_{\beta}\psi_{x_1}^\prime \mathbf{M}\nabla_{\beta}\psi_{x_1}=\nabla_{\beta}\psi_{x_1}^\prime \mathbf{M}^2\nabla_{\beta}\psi_{x_1}= \left(m^{proj}\right)^\prime m^{proj}=0$. Therefore, $m^{proj}=0$.} $\nabla_{\mathbf{H}_{x_1}} \psi_{x_1}^\prime m^{proj}=0$ corresponds to the moment condition that
\begin{equation}
\sum_{j=1}^J m_j^{proj} p_j\left(\alpha_k\right)=E\Big[m^{proj}\left(X_{i}^{1:T},Y_{i}^{1:T}\right)\:|\:X_{i1}=x_1,\alpha_k\Big]=0,\quad \forall \alpha_k\in \mathcal{A}\label{eq:moment_condition}
\end{equation}
where $m^{proj}_j$ denotes the value of the moment function at the $j$ sample paths. Thus, identification of $\beta$ requires the existence of nonzero moment function $\phi$ satisfying the condition (\ref{eq:identification_moment_require}) and (\ref{eq:moment_condition}).

Consider an arbitrary moment function $m$. Under the model specification \ref{eq:y_specification_simple} and the finite support of $\alpha_i$, the equation \eqref{eq:moment_condition} can be represented with matrix form,
\begin{equation}\label{eq:prob_matrix}
\nabla_{\mathbf{H}_{x_1}} \psi_{x_1}^\prime m=
\underbrace{
\begin{pmatrix}
  p_1\left(\alpha_1\right)  & \cdots & p_J\left(\alpha_1\right) \\
  \vdots  & \ddots & \vdots \\
  p_1\left(\alpha_K\right)  & \cdots & p_J\left(\alpha_K\right)
\end{pmatrix}
}_{\nabla_{\mathbf{H}_{x_1}} \psi_{x_1}^\prime=\mathbf{P}}
\begin{pmatrix}
  m_1\\
  \vdots \\
  m_J
\end{pmatrix}
=0.
\end{equation}
If the rank of the matrix $\mathbf{P}$ satisfies
$$
\rank\left(\mathbf{P}\right)<\min\left(|\mathcal{A}|,J\right)=J,
$$
there exists a nonzero moment function $m$. The existence of a moment function is equivalent to identifying the linear independence of the path probabilities among the columns in $\mathbf{P}$:
$$
\sum_{j=1}^J m_j p_j\left(\alpha_k\right)=0,\quad\forall\alpha_k\in\mathcal{A}\quad \Longrightarrow\quad m_j =0,\quad\forall j=1,\hdots,J.
$$
The dependence across the sample paths is the value of the moment function at $j$ sample paths, $m_j$. Consequently, if all path probabilities are linearly independent, there does not exist a nonzero moment function.
 
When $T\geq3$, under Markovian feedback process, there always exist linearly dependent probabilities of the sample path $j,w$ such that
$$
p_j\left(\alpha_k\right)=p_w\left(\alpha_k\right), \quad\forall \alpha_k\in\mathcal{A}.
$$
The following two sample paths are examples of the identical probabilities when $T=3$:
$$
\begin{array}{ccccc}
&(X_{i1},Y_{i1}) &(X_{i2},Y_{i2})&(X_{i3},Y_{i3}) \\\hline\hline
A:& (1,0) &(1,1)&(1,1) \\
B: & (1,1) &(1,0)&(1,1)
\end{array}.
$$
Probabilities of the path $A$ and $B$ are
\begin{equation*}
p_A\left(\alpha_k\right)=p_B\left(\alpha_k\right)
=\exp{\left(2\beta\right)}
A_k^{2}\Big\{1+\exp{\left(\beta\right)A_k\Big\}^{-2}\Big\{1+A_k}\Big\}^{-1} G_{11}\left(\alpha_k\right)G_{10}\left(\alpha_k\right),\quad \forall \alpha_k \in\mathcal{A}.
\end{equation*}
Therefore, there exists a moment function such that for some $j$ and $w$, $m_j+m_w=0$. However, it does not mean that the function $m$ is a projection function satisfying the condition (\ref{eq:identification_moment_require}). The moment function $m$ does not contain any information about $\beta$. 

To distinguish the dependence which is a non-constant function of $\beta$ from the dependence due to the identical probabilities, which are not informative about the parameter, define the former as non-trivial dependence and the latter as trivial dependence.
\begin{definition}
Path probabilities $p_j\left(\alpha_k\right)$ and $p_w\left(\alpha_k\right)$ are \textbf{trivially dependent} if 
$$
p_j\left(\alpha_k\right)=p_w\left(\alpha_k\right),\quad \forall \alpha_k\in\mathcal{A}.
$$
\end{definition}
\begin{definition}
Path probabilities are \textbf{non-trivially dependent} if
$$
\sum_w m_w p_w\left(\alpha_k\right)=0,\quad \forall\alpha_k\in\mathcal{A}
$$
where $m_w$ is a non-constant function of $\beta$ for some $w$.
\end{definition}

In addition to the existence of the moment function satisfying the condition (\ref{eq:moment_condition}), to identify $\beta$ under the Markovian feedback process, there should be a non-trivial moment function $m$ in the sense that the function $m$ is a non-constant function of $\beta$.

From the proposition \ref{pp:setup1}, if the sufficient statistics $\mathbf{S}$, 
$$
\mathbf{S}\left(X_i^{1:T},Y_{i}^{1:T}\right)=\left(\sum_{t=1}^T Y_{it},\left(\sum_{t=2}^T\mathbbm{1}\{X_{it-1}=x_1,Y_{it-1}=y,X_{it}=x_2\}\right)_{x_1,x_2,y\in\{0,1\}}\right)
$$ 
are identical among the paths given the initial condition, the paths have an identical value of $\sum_{t=1}^T X_{it}Y_{it}$. Under the simple specification \ref{eq:y_specification_simple}, the paths have identical probability. Therefore, if there exist paths such that the probabilities of the paths are non-trivially linearly dependent, at least one of the statistics should be different.
\begin{corollary}\label{cr:identification_condition}
If there are non-trivially linearly dependent probabilities of the paths, at least one value of $\left(\sum_{t=1}^T Y_{it}, \left(\sum_{t=2}^T\mathbbm{1}\{X_{it-1}=x_1,Y_{it-1}=y,X_{it}=x_2\}\right)_{x_1,x_2,y\in\{0,1\}}\right)$ is different among some paths.
\end{corollary}

Suppose there are $w_s$ sample paths in $\mathcal{D}_s$. By the proposition \ref{pp:setup1}, $p_w=p_{s}$ holds for the path $w\in \mathcal{D}_{s}$. A linear combination of the paths is
\begin{equation}
\sum_{j=1}^J m_jp_j\left(\alpha_k\right)=\sum_{s\in \mathcal{S}}\sum_{w=1}^{w_s} m_{w} p_{w}\left(\alpha_k\right)=\sum_{s\in \mathcal{S}}\underbrace{\left(\sum_{w=1}^{w_s} m_{w}\right)}_{=m_{s}}p_{s}\left(\alpha_k\right),\quad\forall \alpha_k\in\mathcal{A}.\label{eq:linear_comb} 
\end{equation}
Consequently, the trivial dependence across probabilities of the paths in $\mathcal{D}_s$ can be reduced to one representative probability $p_{s}\left(\alpha_k\right)$ with $m_{s}$. In the subsequent lemmas and proposition, the representative probability $p_{s}\left(\alpha_k\right)$ will be used to avoid trivial dependent path probabilities.

The following lemmas restrict the sets of the paths which have non-trivially linearly dependent probabilities. The lemmas state that the paths should have the same polynomial degrees in terms of the nuisance parameters to have the dependence free of the nuisance parameters. 
\begin{restatable}{lemma}{Lemidenticalsumy} \label{lm:identical_sum_y}
If there are non-trivial linear dependence among probabilities of the paths, the paths have identical values of the statistics $\sum_{t=1}^T Y_{it}$.
\end{restatable}
For the proof of Lemma \ref{lm:identical_sum_y}, see Appendix \ref{ap:proof}. Lemma \ref{lm:identical_sum_y} states that if there exists non-trivial dependence free of the nuisance parameters, the polynomial degrees of $A_i$ among the linearly dependent path probabilities should be identical, $n_y^j=n_y$.

Similarly, the linearly dependent probabilities of the sample paths have identical degrees in terms of $G_{xy}\left(\alpha_k\right)$ for all $x,y\in\{0,1\}$.
\begin{restatable}{lemma}{LemidenticalorderG}
\label{lm:identical_order_G}
If there is non-trivial linear dependence among probabilities of the paths, the paths have identical values of the statistics 
$$
\sum_{t=2}^T\mathbbm{1}\{X_{it-1}=x_1,Y_{it-1}=y\}=\sum_{t=2}^T\mathbbm{1}\{X_{it-1}=x_1,Y_{it-1}=y,X_{it}=0\}+\sum_{t=2}^T\mathbbm{1}\{X_{it-1}=x_1,Y_{it-1}=y,X_{it}=1\}
$$ for all $x,y\in\{0,1\}$.
\end{restatable}
For the proof of Lemma \ref{lm:identical_order_G}, see Appendix \ref{ap:proof}. Lemma \ref{lm:identical_order_G} states that if the path probabilities have nontrivial linear dependence, the paths have identical $n_{x,y}$ such that
$$
n_{x,y}=n_{x,y}^j=n_{x,1,y}^{j}+n_{x,0,y}^{j},\quad \forall x,y\in\{0,1\}
$$
to match the polynomial degree in $G_{xy}\left(\alpha_k\right)$ across the path probabilities which are non-trivially linearly dependent. 

Under Corollary \ref{cr:identification_condition}, Lemma \ref{lm:identical_sum_y}, and Lemma \ref{lm:identical_order_G}, Lemma \ref{lm:no_nontrivial_depen} establishes the nonexistence of nontrivial linear dependence among path probabilities. 
\begin{restatable}{lemma}{Lemnonotrivial}
\label{lm:no_nontrivial_depen}
There is no non-trivial linear dependence among probabilities of the paths.
\end{restatable}
For the proof of Lemma \ref{lm:no_nontrivial_depen}, see Appendix \ref{ap:proof}. Consequently, according to Lemma \ref{lm:identical_sum_y},\ref{lm:identical_order_G}, and \ref{lm:no_nontrivial_depen}, there is only trivial dependence among the paths and $m$ is the moment function reflecting trivial dependence. Since there is no non-trivial dependence, from the linear combination of the path probabilities \ref{eq:linear_comb}, all $m_s$ are zero.
$$
\sum_{s\in \mathcal{S}}m _{s}p_{s}\left(\alpha_k\right)=0,\quad\forall \alpha_k\in\mathcal{A}\quad \Longrightarrow \quad 
m_{s}=\sum_{w=1}^{w_s} m_{w}=0,\quad \forall s\in \mathcal{S}
$$
It implies that $\nabla_{\beta} \psi_{x_1}^\prime m=0$ since
$$
\nabla_{\beta} \psi_{x_1}^\prime m=\sum_{\alpha_k \in \mathcal{A}}H_{x_1}\left(\alpha_k\right)\sum_{s\in \mathcal{S}}\sum_{w=1}^{w_s} m_{w} \frac{\partial p_{w}\left(\alpha_k\right)}{\partial \beta}=\sum_{\alpha_k \in \mathcal{A}}H_{x_1}\left(\alpha_k\right)\sum_{s\in \mathcal{S}}\underbrace{\left(\sum_{w=1}^{w_s} m_{w}\right)}_{=m_s=0}\frac{\partial p_{s}\left(\alpha_k\right)}{\partial \beta}=0, 
$$
where second equality holds because $p_w=p_s$ for all paths in $\mathcal{D}_s$ and third equality holds due to $m_s=0$ for all $s\in\mathcal{S}$. $m$ is not a projection function satisfying $\nabla_{\beta} \psi_{x_1}^\prime m\neq0$. Therefore, although there exist  moment functions, the functions do not satisfy the conditions for the projection function. Since the projection function does not exist, $\beta$ is not identified regardless of the time period $T$ in the simple model specification \ref{eq:y_specification_simple}.

\begin{restatable}{proposition}{propnonidentification}
\label{pp:non_identification}
Consider the following dynamic logit model specification under Assumptions \ref{as:compact_parameter} - \ref{as:markov_setup1}, and \ref{as:regular_point}:
\begin{multline*}
P\left(X_i^{2:T}=x^{2:T},Y_{i}^{1:T}=y^{1:T}\:|\:X_{i1}=x_1,\alpha_i\:;\:\beta\right)
\\=\exp{\left(\beta \sum_{t=1}^T x_{t}y_{t}\right)}
\frac{\exp\left(\alpha_i\sum_{t=1}^T y_{t}\right)}{\prod_{t=1}^T\Big\{1+\exp{\left(\beta x_{t}+\alpha_i\right)}\Big\}} \prod_{t=2}^T G\left(x_t\:|\:x_{t-1},y_{t-1}\:;\:\alpha_i\right).
\end{multline*}
For any $T$, identification of $\beta$ fails. 
\end{restatable}

Proposition \ref{pp:non_identification} states that a vector of observed probabilities does not provide sufficient information to identify $\beta$. In addition to the observed probabilities, the model may imply additional valid moment conditions, such as those arising from the sequential exogeneity of $X_{it}$, although these are challenging to exploit in the nonlinear structure of the logit specification. 

Consistent with the non-identification of $\beta$ under the logistic specification with an unrestricted feedback process in \textcite{bonhomme_identification_2023}, Proposition \ref{pp:non_identification} shows that $\beta$ is not identified even under a time-homogeneous first-order Markovian feedback process, which is more restrictive. This non-identification result necessitates additional restrictions to identify $\beta$ under the logistic specification, which will be investigated in Section \ref{sc:additional_assumption}.
\begin{remark}
    \textcite{bonhomme_identification_2023} also investigates identification under additional restrictions on the feedback process. One of the restrictions is a first-order Markovian feedback process which may vary across time periods $t$,
    $$
    G_i^t\left(x_{t}\:|\:x^{1:t-1},y^{1:t-1}\right)=G_i^t\left(x_{t}\:|\:x_{t-1},y_{t-1}\right).
    $$
    \textcite{bonhomme_identification_2023} argues that identification under this restriction requires methods beyond those used in the paper.
    Proposition \ref{pp:non_identification} establishes a non-identification result in a panel logit model under the more restrictive case in which the feedback is time-invariant,
    $$
    G_i^t\left(x_{t}\:|\:x_{t-1},y_{t-1}\right)=G_i\left(x_{t}\:|\:x_{t-1},y_{t-1}\right),\quad \forall t=2,\hdots,T.
    $$
    Since the time-invariant specification is nested within the class of the time-varying first-order feedback process, our result implies non-identification in a panel logit model for the broader class.
\end{remark}

\section{Identification under Additional Assumptions}\label{sc:additional_assumption}
Non-identification result in Proposition \ref{pp:non_identification} implies that additional assumptions are required to identify $\beta$. This section considers two additional assumptions for the identification while holding heterogeneity of the feedback process across individuals. First assumption is imposed on the feedback process of the covariates. If the feedback process depends on $Y_{it-1}$ only, $\beta$ is identified via CMLE when $T\geq2$. Second assumption is imposed on the initial condition. If initial condition and the unobserved heterogeneity are independent, it provides source of the identification of $\beta$ via CMLE when $T\geq3$. This section also briefly investigates analogous identification result with multiple sequentially exogenous variable and discrete strict exogenous variables.
\subsection{Assumption on the Feedback Process}\label{ssc:markov_setup2}
Consider logit model specification \ref{eq:y_specification} with a different feedback process where $\mathbf{G}_i$ depends only on $Y_{it-1}$.
\begin{assumption}[Markovian feedback process of $X_{it}$]\label{as:markov_setup2}
Discrete $X_{it}$, defined on a finite support $\mathcal{X}$, follows a first-order Markov process. $\mathbf{G}_i$ is a Markov kernel from $\{0,1\}$ to the interior of the unit simplex on $\mathcal{X}$, 
$$\mathbf{G}_i:\{0,1\}\rightarrow \Delta_{+}^\mathcal{X}$$
with entries
$G_i\left(x\:|\:y\right)=P\left(X_{it}=x\:|\:Y_{it-1}=y\right)$
for all $y\in\{0,1\}$ and $x\in\mathcal{X}$. $\mathcal{G}$ is the collection of all Markov kernel $\mathbf{G}_i$.
\end{assumption}
Under Assumptions \ref{as:stationary} - \ref{as:iida}, and \ref{as:markov_setup2}, a joint probability of $\left(X_i^{1:T},Y_i^{1:T}\right)$ at the path $\left(x^{1:T},y^{1:T}\right)$ given the initial condition $Y_{i0}=y_0$ is
\begin{multline*}
P\left(X_i^{1:T}=x^{1:T},Y_{i}^{1:T}=y^{1:T}\:|\:Y_{i0}=y_0,\alpha_i,\mathbf{G}_i\:;\:\theta\right)=\underbrace{\exp{\left(\rho \sum_{t=1}^T y_{t-1}y_{t}+\beta \sum_{t=1}^T x_{t}y_{t}\right)}}_{f_\theta\left(x^{1:T},y^{0:T}\right)}\\
\times\underbrace{\frac{\exp\left(\alpha_i\sum_{t=1}^T y_{t}\right)}{\prod_{t=1}^T\Big\{1+\exp{\left(\rho y_{t-1}+\beta x_{t}+\alpha_i\right)}\Big\}}}_{f_{\alpha_i}\left(s^\alpha\right)}\underbrace{\prod_{y\in\{0,1\}}\prod_{x\in\mathcal{X}} G_i\left(x\:|\:y\right)^{\sum_{t=1}^{T} \mathbbm{1}\{y_{t-1}=y,x_{t}=x\}}}_{f_{\mathbf{G}_i}\left(s^\mathbf{G}\right)}.\label{eq:logit_setup2}
\end{multline*}
Analogous to the sufficient statistics \ref{eq:suff_model1} for the first model specification \ref{eq:y_specification}, associated sufficient statistics for $\alpha_i$ and $\mathbf{G}_i$ are, respectively,
\begin{align*}
\mathbf{S}^\alpha\left(X_{i}^{1:T},Y_{i}^{0:T}\right)&=\left(\sum_{t=1}^T Y_{it},\left(\sum_{t=1}^{T} \mathbbm{1}\{Y_{it-1}=y,X_{it}=x\} \right)_{y\in\{0,1\}, x\in\mathcal{X}}\right),\\
\mathbf{S}^{\mathbf{G}}\left(X_{i}^{1:T},Y_{i}^{0:T}\right)&=\left(\sum_{t=1}^{T} \mathbbm{1}\{Y_{it-1}=y,X_{it}=x\} \right)_{y\in\{0,1\}, x\in\mathcal{X}}.
\end{align*}
Notice that sufficient statistics for transition of $X$ in $f_{\mathbf{G}_i}$ is different from those in $f_{\mathbf{G}_i}$ from equation (\ref{eq:setup1_f3}) since $\mathbf{G}_i$ does not depend on $X_{it-1}$. Therefore, sufficient statistics $\mathbf{S}$ for the nuisance parameters are given below,
\begin{align*}
\mathbf{S}\left(X_{i}^{1:T},Y_{i}^{0:T}\right)&=\left(\mathbf{S}^\alpha\left(X_{i}^{1:T},Y_{i}^{0:T}\right),\mathbf{S}^\mathbf{G}\left(X_{i}^{1:T},Y_{i}^{0:T}\right)\right)\\
&=\left(\sum_{t=1}^TY_{it}, \left(\sum_{t=1}^{T} \mathbbm{1}\{Y_{it-1}=y,X_{it}=x\}\right)_{ y\in\{0,1\},x\in\mathcal{X}} \right).
\end{align*}
The statistics impose weaker constraints on possible paths $(X_{i}^{1:T},Y_{i}^{1:T})$ conditional on $\mathbf{S}$ than the statistics \ref{eq:suff_model1}. These weaker restrictions allow for the identification of $\beta$ and $\rho$, which is shown in Proposition \ref{pp:setup2}.

Similar to the set \ref{eq:history_set_setup1}, let $\mathcal{D}_s$ denote the set of the paths of $(x^{2:T},y^{1:T})$ such that $\mathbf{S}=s$ be
\begin{equation}
\mathcal{D}_s=\Big\{\left(x^{1:T},y^{1:T}\right):\mathbf{S}\left(x^{1:T},y^{0:T}\right)=s\Big\}, \quad \forall s\in\mathcal{S}.\label{eq:history_set_setup2}
\end{equation}
The probability of $\left(X_i^{1:T},Y_{i}^{1:T}\right)$ at $\left(x^{1:T},y^{1:T}\right)$ conditional on $\mathbf{S}=s$ and the initial condition $Y_{i0}=y_0$ is
\begin{multline}
P\left(X_i^{1:T}=x^{1:T},Y_{i}^{1:T}=y^{1:T}\:|\:\left(X_i^{1:T},Y_i^{1:T}\right)\in \mathcal{D}_s,Y_{i0}=y_0\:;\:\theta\right)\\
=\frac{\exp{\left(\rho \sum_{t=1}^T y_{t-1} y_{t}+\beta\sum_{t=1}^T x_{t}y_{t}\right)}}{\sum_{\left(\tilde{x}^{1:T},\tilde{y}^{1:T}\right)\in \mathcal{D}_s}\exp{\left(\rho \sum_{t=1}^T \tilde{y}_{t-1} \tilde{y}_{t}+\beta\sum_{t=1}^T \tilde{x}_{t} \tilde{y}_{t}\right)}},\quad \forall s \in\mathcal{S}.\label{eq:condi_prob_as1}
\end{multline}
where $\tilde{x}_{t},\tilde{y}_{t}$ denotes $x_{t},y_{t}$ in $(\tilde{x}^{1:T},\tilde{y}^{1:T})$ and $\tilde{y}_{0}=y_0$.

Under the condition (\ref{eq:identifying_stat}), Proposition \ref{pp:setup2} establishes identification result via conditional likelihood.
\begin{restatable}{proposition}{propsetupyonly}\label{pp:setup2}
Consider following dynamic logit model specification under Assumptions \ref{as:compact_parameter} - \ref{as:iida}, and \ref{as:markov_setup2}:
\begin{multline*}
P\left(X_i^{1:T}=x^{1:T},Y_{i}^{1:T}=y^{1:T}\:|\:Y_{i0}=y_0,\alpha_i,\mathbf{G}_i\:;\:\theta\right)\\
=\exp{\left(\rho \sum_{t=1}^T y_{t-1}y_{t}+\beta \sum_{t=1}^T x_{t}y_{t}\right)}
\times\frac{\exp\left(\alpha_i\sum_{t=1}^T y_{t}\right)}{\prod_{t=1}^T\Big\{1+\exp{\left(\rho y_{t-1}+\beta x_{t}+\alpha_i\right)}\Big\}}\prod_{t=1}^T G_i\left(x_t\:|\:y_{t-1}\right).
\end{multline*}
For $T\geq2$, $\beta$ is identified via conditional likelihood. For $T\geq3$, $\rho$ is identified via conditional likelihood.
\end{restatable}
For the proof of Proposition \ref{pp:setup2}, see Appendix \ref{ap:proof}. As an example for the identification of $\beta$ when $T=2$, consider next example:
$$
\begin{array}{cccc}
&Y_{i0}&(X_{i1},Y_{i1}) &(X_{i2},Y_{i2})\\\hline\hline
A:&1 & (1,1) &(0,0) \\
B:&1& (0,1) &(1,0)\\
\end{array}.
$$
Path $A$ and $B$ have the same value of the statistics $\mathbf{S}$: There is an identical number of occurrences of feedback $(Y_{it-1},X_{it})$ over periods, $(1,1)$, $(1,0)$, in $A$ and $B$ as well as $\sum_{t=1}^T Y_{it}^A=\sum_{t=1}^T Y_{it}^B=1$. However, $A$ and $B$ have different values in $\sum_{t=1}^T X_{it}Y_{it}$,
$$
\sum_{t=1}^T X_{it}^A Y_{it}^A=1, \sum_{t=1}^T X_{it}^B Y_{it}^B=0.
$$ Therefore, there exists $\mathbf{S}=s$ such that the distributions of $\sum_{t=1}^T X_{it} Y_{it}$ are not degenerate once conditional on $\mathbf{S}=s$. Condition (\ref{eq:identifying_stat}) holds and $\beta$ is identified when $T=2$.

To illustrate identification of $\rho$ when $T=3$, consider the following three paths $A,B$ and $C$ below
$$
\begin{array}{ccccc}
&Y_{i0}&(X_{i1},Y_{i1}) &(X_{i2},Y_{i2})&(X_{i3},Y_{i3}) \\\hline\hline
A:&0 & (1,0) &(0,1)&(1,1) \\
B:&0 & (0,1) &(1,0)&(1,1)\\
C:&0 & (1,1) &(1,0)&(0,1)
\end{array}
$$
where $A,B$ and $C$ have the same values of sufficient statistics: They have the same transitions of $(Y_{it-1},X_{it})$ over time: $(0,1)$, $(0,0)$, $(1,1)$.
Also, $\sum_{t=1}^T Y^A_{it}=\sum_{t=1}^T Y^B_{it}=\sum_{t=1}^T Y^C_{it}=2.$
In terms of $\sum_{t=1}^T Y_{it-1} Y_{it}$, $A,B$ and $C$ have different values:
$$
\sum_{t=1}^T Y_{it-1}^A Y_{it}^A=1,\sum_{t=1}^T Y_{it-1}^B Y_{it}^B=\sum_{t=1}^T Y_{it-1}^C Y_{it}^C=0.
$$ Thus, the condition (\ref{eq:identifying_stat}) is satisfied and $\rho$ is identified when $T=3$.

Limited dependence of the feedback process of $X_{it}$ on $Y_{it-1}$ from the Assumption \ref{as:markov_setup2} enables the identification of $\beta$ since the process in each period is independent of those in the other period. The process \ref{as:markov_setup1} across periods is not independent because $\mathbf{G}_i$ depends on $X_{it-1}$ and $Y_{it-1}$. Therefore, feedback in the previous period directly limits the possible feedback in subsequent periods, which restricts possible paths and the corresponding value of $\sum_{t=1}^T X_{it}Y_{it}$ conditional on the sufficient statistics.

\begin{remark}
The identification result of $\beta$ in Proposition \ref{pp:setup2} under Assumption \ref{as:markov_setup2} is consistent to the identification result in \textcite{Honore_Moment_functional_2024}. In the supplementary materials of \textcite{Honore_Moment_functional_2024}, the authors state that the moment method in the main text of the paper can be extended to predetermined $X_{it}$ in the sense that $X_{it}$ depends on $Y_{it-1}$ which aligns with Assumption \ref{as:markov_setup2},
$$
X_{it}=h\left(Y_{it-1},\tilde{X}_{it}\right)
$$
where $\tilde{X}_{it}$ is an unobserved exogenous variable. Proposition \ref{pp:setup2} shows that conditional likelihood is also a plausible approach of for identifying $\beta$ under the restricted dependence of $X_{it}$ described in Assumption \ref{as:markov_setup2}.
\end{remark}

\subsection{Assumption on the Initial Condition}\label{ssc:initial_condition}
Assumption \ref{as:markov_setup2} restricts the feedback process in that it depends only on $Y_{it-1}$, not $X_{it-1}$. An alternative way to identify $\beta$ while maintaining the first-order Markovian feedback process assumption \ref{as:markov_setup1} is to impose a restriction on the initial condition. Consider panel data $\left(X_{i}^{0:T},Y_{i}^{0:T}\right)$ over $i=1,\hdots,N$ individuals where the initial condition is $I_i=\left(X_{i0},Y_{i0}\right)$ which is observed from the data. Assume that the distribution of the initial condition is independent of the nuisance parameters.
\begin{assumption}[Independent and identically distributed $\alpha_i$]\label{as:iida_restriction}
$\alpha_i$ is a sequence of independent identically distributed random variables with the probability density function $H\left(\alpha\right)$ and support $\mathcal{A}$. $\alpha_i$ and $\mathbf{G}_i$ are independent of $I_i=\left(X_{i0},Y_{i0}\right)$.
\end{assumption}

Let $p$ denote the vector of the initial condition probabilities, $p=\left(p_{xy}\right)_{x\in\mathcal{X},y\in\{0,1\}}^\prime$ where $$
p_{xy}=P\left(X_{i0}=x,Y_{i0}=y\right)$$
and let $\mathcal{P}$ be the admissible set of probability vectors. We also assume that the initial condition probabilities are uniformly bounded away from zero.
\begin{assumption}[Uniform lower bound on initial condition probabilities]\label{as:initial_positive}
The initial condition probabilities are uniformly bounded away from zero. That is, there exists $\delta>0$ such that for all $p\in \mathcal{P}$,
$$
p_{xy}\geq\delta,\quad\forall\left(x,y\right)\in \mathcal{X}\times \{0,1\}.
$$
\end{assumption}
With Assumption \ref{as:markov_setup1} and \ref{as:initial_positive}, every path of $\left(X_{i}^{0:T},Y_{i}^{0:T}\right)$ is attainable with positive probability. The conditional likelihood of $\left(X_{i}^{0:T},Y_{i}^{0:T}\right)$ at $\left(x^{0:T},y^{0:T}\right)$ is
\begin{align*}
&P\left(X_i^{0:T}=x^{0:T},Y_{i}^{0:T}=y^{0:T}\:|\:\alpha_i,\mathbf{G}_i\:;\:\theta\right)\\
&=P\left(X_i^{1:T}=x^{1:T},Y_{i}^{1:T}=y^{1:T}\:|\:X_{i0}=x_0,Y_{i0}=y_0,\alpha_i,\mathbf{G}_i\:;\:\theta\right)P\left(X_{i0}=x_0,Y_{i0}=y_0\:|\:\alpha_i,\mathbf{G}_i\right)\\
&=P\left(X_i^{1:T}=x^{1:T},Y_{i}^{1:T}=y^{1:T}\:|\:X_{i0}=x_0,Y_{i0}=y_0,\alpha_i,\mathbf{G}_i\:;\:\theta\right)P\left(X_{i0}=x_0,Y_{i0}=y_0\right)
\end{align*}
where last equality holds due to Assumption \ref{as:iida_restriction}. Analogous to the factorization \ref{eq:setup1_factorization}, the likelihood is decomposed,
\begin{multline*}
P\left(X_i^{0:T}=x^{0:T},Y_{i}^{0:T}=y^{0:T}\:|\:\alpha_i,\mathbf{G}_i\:;\:\theta\right)\\
=\underbrace{p_{x_0,y_0}\exp{\left(\rho \sum_{t=1}^T y_{t-1}y_{t}+\beta \sum_{t=1}^T x_{t}y_{t}\right)}}_{f_{\theta}\left(x^{0:T},y^{0:T}\right)}
\underbrace{\frac{\exp\left(\alpha_i\sum_{t=1}^T y_{t}\right)}{\prod_{t=1}^T\Big\{1+\exp{\left(\rho y_{t-1}+\beta x_{t}+\alpha_i\right)}\Big\}}}_{f_{\alpha_i}\left(x^{1:T},y^{0:T}\right)}\underbrace{\prod_{t=1}^T G_i\left(x_t\:|\:x_{t-1},y_{t-1}\right)}_{f_{\mathbf{G}_i}\left(x^{0:T},y^{0:T-1}\right)}.
\end{multline*}
Notice that $X_{i1}$ is not initial condition and has feedback process given $\left(X_{i0},Y_{i0}\right)$. Corresponding sufficient statistics are
\begin{align*}
\mathbf{S}\left(X_i^{0:T},Y_{i}^{0:T}\right)&=\left(\mathbf{S}^\alpha\left(X_i^{1:T},Y_{i}^{0:T}\right),\mathbf{S}^\mathbf{G}\left(X_i^{0:T},Y_{i}^{0:T-1}\right)\right)\nonumber\\
&=\left(\sum_{t=1}^TY_{it},\left(\sum_{t=1}^{T} \mathbbm{1}\{X_{it-1}=x_1,Y_{it-1}=y,X_{it}=x_2\}\right)_{x_1,x_2\in\mathcal{X},y\in\{0,1\}}\right).
\end{align*}
The possible sample paths given sufficient statistics are
\begin{equation*}
\mathcal{D}_s=\Big\{\left(x^{0:T},y^{0:T}\right):\mathbf{S}\left(x^{0:T},y^{0:T}\right)=s\Big\}.
\end{equation*}
The probability of $\left(X_i^{0:T},Y_{i}^{0:T}\right)$ at $\left(x^{0:T},y^{0:T}\right)$ conditional on $\mathbf{S}=s$ is
\begin{align}
&P\left(X_i^{0:T}=x^{0:T},Y_{i}^{0:T}=y^{0:T}\:|\:\left(X_i^{0:T},Y_i^{0:T}\right)\in \mathcal{D}_s\:;\:\theta\right)\nonumber\\
&=\frac{p_{x_0,y_0}\exp{\left(\rho \sum_{t=1}^T y_{t-1} y_{t}+\beta\sum_{t=1}^T x_{t}y_{t}\right)}}{\sum_{\left(\tilde{x}^{0:T},\tilde{y}^{0:T}\right)\in \mathcal{D}_s}p_{\tilde{x}_0,\tilde{y}_0}\exp{\left(\rho \sum_{t=1}^T \tilde{y}_{t-1} \tilde{y}_{t}+\beta\sum_{t=1}^T \tilde{x}_{t} \tilde{y}_{t}\right)}},\quad \forall s \in\mathcal{S}.\label{eq:condi_prob_as2}
\end{align}
Since the conditional probability does not condition on the initial condition, paths in $\mathcal{D}_s$ may have different initial conditions.

By Assumption \ref{as:iida_restriction}, the initial condition probabilities do not depend on $\alpha_i$ and $\mathbf{G}_i$, and they are estimated from the data. Once the initial condition probabilities are consistently estimated from the data, such as a frequency estimator, they can be used for identification of the parameters via pseudo CMLE. If the initial condition probabilities are instead jointly estimated via CMLE with $\beta$, $\beta$ is not identified because the variation in $\sum_{t=1}^T X_{it}Y_{it}$ can be absorbed by the initial condition probabilities. Thus, when identifying $\beta$, the initial condition probabilities should be estimated separately.

\begin{restatable}{proposition}{propsetupinitial}\label{pp:setup3}
Consider the following dynamic logit model specification under Assumptions \ref{as:compact_parameter}, \ref{as:stationary}, \ref{as:markov_setup1}, \ref{as:iida_restriction}, and \ref{as:initial_positive}:
\begin{multline*}
P\left(X_i^{0:T}=x^{0:T},Y_{i}^{0:T}=y^{0:T}\:|\:\alpha_i,\mathbf{G}_i\:;\:\theta\right)
=P\left(X_{i0}=x_0,Y_{i0}=y_0\right)\\\times \exp{\left(\rho \sum_{t=1}^T y_{t-1}y_{t}+\beta \sum_{t=1}^T x_{t}y_{t}\right)}
\frac{\exp\left(\alpha_i\sum_{t=1}^T y_{t}\right)}{\prod_{t=1}^T\Big\{1+\exp{\left(\rho y_{t-1}+\beta x_{t}+\alpha_i\right)}\Big\}} \prod_{t=1}^T G_i\left(x_t\:|\:x_{t-1},y_{t-1}\right).
\end{multline*}
For $T\geq2$, $\beta$ is identified via conditional likelihood. For $T\geq3$, $\rho$ is identified via conditional likelihood.
\end{restatable}
For the proof of Proposition \ref{pp:setup3}, see Appendix \ref{ap:proof}. Consider the following two sample paths which have identical value of sufficient statistics when $T=2$:
$$
\begin{array}{ccccc}
&\left(X_{i0},Y_{i0}\right)&(X_{i1},Y_{i1}) &(X_{i2},Y_{i2}) \\\hline\hline
A:&(0,1)& (1,1) &(0,0) \\
B:&(1,1)& (0,1) &(1,0)
\end{array}.
$$
Path $A$ and $B$ have identical $\sum_{t=1}^2 Y_{it}$ and occurrence of feedback: $\sum_{t=1}^2 Y_{it}^A=\sum_{t=1}^2 Y_{it}^B=1$, $G_i\left(1\:|\:0,1\right)$, and $G_i\left(0\:|\:1,1\right)$. On the other hand, $A$ and $B$ do not have identical values of $\sum_{t=1}^2 X_{it}Y_{it}$,
$$
\sum_{t=1}^2 X_{it}^A Y_{it}^A=1, \sum_{t=1}^2 X_{it}^B Y_{it}^B=0.
$$
Thus, $\beta$ is identified via conditional likelihood when $T=2$ since condition (\ref{eq:identifying_stat}) holds. Since $p_{01}$ and $p_{11}$ are identified from the data, the same identification condition (\ref{eq:identifying_stat}) can be applied.

For the identification of $\rho$ when $T=3$, consider the following two sample paths,
$$
\begin{array}{ccccc}
&\left(X_{i0},Y_{i0}\right)&(X_{i1},Y_{i1}) &(X_{i2},Y_{i2}) &(X_{i3},Y_{i3}) \\\hline\hline
A:&(0,0)& (0,0) &(0,1) &(0,1)\\
B:&(0,0)& (0,1) &(0,0) &(0,1)
\end{array}.
$$
Each path has identical feedback processes, $G_i\left(0\:|\:0,0\right)$, $G_i\left(0\:|\:0,0\right)$, and $G_i\left(0\:|\:1,0\right)$ and identical $\sum_{t=1}^3 Y_{it}^A=\sum_{t=1}^3 Y_{it}^B=2$. $\sum_{t=1}^3 Y_{it-1} Y_{it}$ of path A and B are
$$
\sum_{t=1}^3 Y_{it-1}^A Y_{it}^A=1, \sum_{t=1}^3 Y_{it-1}^B Y_{it}^B=0.
$$
Consequently, there exist at least two sample paths satisfying the condition (\ref{eq:identifying_stat}) and $p_{00}$ is identified from the data, $\rho$ is identified when $T=3$, which can be generalized to the case with longer time periods.

Proposition \ref{pp:setup2} and \ref{pp:setup3} establish identification of $\beta$ under Assumptions \ref{as:markov_setup2} and \ref{as:iida_restriction}, respectively, while maintaining sequential exogeneity of $X_{it}$ and without imposing any parametric restriction on the feedback process.


\subsection{Extension}\label{ssc:strict_exo}
Identification in Propositions \ref{pp:setup2} and Proposition \ref{pp:setup3} extends to model specifications with multiple sequentially exogenous variables and strictly exogenous covariates defined on finite supports. Consider the following model specification:
\begin{equation*}
Y_{it}=\mathbbm{1}\Big\{\rho Y_{it-1}+ X_{it}^\prime \beta+Z_{it}^\prime \gamma+\alpha_i\geq u_{it}\Big\},\quad \forall i= 1,\hdots,N,\quad \forall t=1,\hdots,T,
\end{equation*}
where $X_{it}$ and $Z_{it}$ are vectors of discrete sequentially exogenous and strictly exogenous variables, respectively. The support of $Z_{it}$, denoted by $\mathcal{Z}$, is finite. We further allow that the law of motion of $X_{it}$ to depend on $Z_{it}$. The rest of the identification argument is unchanged. For example, if the conditional distribution of $X_{it}$ depends on $\left(Y_{it-1},Z_{it}\right)$, which is an analogous extension of Assumption \ref{as:markov_setup2}, sufficient statistics for $\alpha_i$ and $\mathbf{G}_i$ are 
\begin{align*}
\mathbf{S}\left(X_{i}^{1:T},Y_{i}^{0:T}\right)=\left(\sum_{t=1}^TY_{it}, \left(\sum_{t=1}^{T} \mathbbm{1}\{Y_{it-1}=y,X_{it}=x,Z_{it}=z\}\right)_{ y\in\{0,1\},x\in\mathcal{X},z\in \mathcal{Z}} \right).
\end{align*}
The resulting conditional probability is
\begin{multline*}
P\left(X_i^{1:T}=x^{1:T},Y_{i}^{1:T}=y^{1:T}\:|\:\left(X_i^{1:T},Y_i^{1:T}\right)\in \mathcal{D}_s,Y_{i0}=y_0,Z_{i}^{1:T}=z^{1:T}\:;\:\theta\right)\\
=\frac{\exp{\left(\rho \sum_{t=1}^T y_{t-1}y_{t}+\sum_{t=1}^T y_t x_{t}^\prime \beta+\sum_{t=1}^T y_t z_{t}^\prime \gamma\right)}}{\sum_{\left(\tilde{x}^{1:T},\tilde{y}^{1:T}\right)\in \mathcal{D}_s}\exp{\left(\rho \sum_{t=1}^T \tilde{y}_{t-1}\tilde{y}_{t}+\sum_{t=1}^T \tilde{y}_t \tilde{x}_{t}^\prime \beta+\sum_{t=1}^T \tilde{y}_t z_{t}^\prime \gamma\right)}},\quad \forall s \in\mathcal{S}.
\end{multline*}
where $\tilde{x}_{t},\tilde{y}_{t}$ denote $x_{t}$ and $y_{t}$ in $(\tilde{x}^{1:T},\tilde{y}^{1:T})$ and $\tilde{y}_{0}=y_0$. Analogously to Proposition \ref{pp:setup2}, the existence of strictly exogenous $Z_{it}$ and multiple sequentially exogenous covariates does not affect the identification result. Identification via conditional likelihood with Assumption \ref{as:markov_setup1} augmented with the strictly exogenous covariates in Section \ref{ssc:initial_condition} is plausible in a similar manner. Aside from the specifications described above, there are many interesting extensions of the identification via conditional likelihood under sequential exogeneity once the covariates are discrete.

\section{Conclusion}\label{sc:conclusion}
This paper investigates identification in dynamic panel logit models with the first-order Markovian feedback process, state dependence, and unobserved heterogeneity. We provide sufficient statistics for unobserved heterogeneity and the feedback process when the process of the covariate depends on the lagged dependent variable and the lagged covariate. These sufficient statistics can be generalized to settings that include discrete, strictly exogenous covariates, which may be useful for future research.

We also analyze the failure of identification, which necessitates additional identifying restrictions. Furthermore, we suggest two constraints for the identification of the sequentially exogenous variable without imposing parametric structure on the process: one is to impose that the covariate depends only on the lagged dependent variable and not on lagged values of the covariate, and the other is to impose a strictly exogenous initial condition, both of which allow the heterogeneous feedback processes.

In this paper, we derive identification by introducing sufficient statistics that rely on the discrete nature of variables. However, conditional maximum likelihood estimation is not constrained to discrete variables. The logic described in the paper may be extended to a continuous covariate. Furthermore, sufficient statistics for the feedback process do not depend on the logit specification, implying that there can be other model specifications which may allow identification under such feedback processes. We leave these for future research.

\printbibliography

\clearpage

\appendix
\section*{Appendix}

\section{Uniqueness of the Maximizer}\label{ap:many_proof}
\subsection*{Uniqueness of the Maximizer of the Sample Objective Function of the CMLE}
Let $W_{it}$ and $w_{it}$ denote the corresponding regressors for the identified parameters and realized values of the regressors, respectively, with $W_{it}=\left(Y_{it-1},X_{it}\right)^\prime$. Denote the log likelihood at the path $d_i$ conditional on $\mathbf{S}_i=s_i$ as $\ln \phi_\theta\left(d_i\:|\:s_i\right)$.
$$
\ln \phi_\theta\left(d_i\:|\:s_i\right)=\sum_{t=1}^Ty_{it}w_{it}^\prime\theta-\ln\left(\sum_{\tilde{d}_i\in \mathcal{D}_{s_i}}\exp\left(\sum_{t=1}^T\tilde{y}_{it}\tilde{w}_{it}^\prime\theta\right)\right).
$$
Since $\Theta$ is assumed to be compact by Assumption \ref{as:compact_parameter} and the likelihood (\ref{eq:conditional_likelihood}) is continuous in $\theta$, there exists at least one maximizer of the conditional likelihood. 

The log likelihood conditional on $\mathbf{S}_i=s_i$ is concave. Second derivative of $\ln \phi_\theta\left(d_i\:|\:s_i\right)$,
\begin{equation*}
-\frac{\sum_{\tilde{d}_i,\bar{d}_i\in \mathcal{D}_{s_i}}\exp\left(\sum_{t=1}^T\tilde{y}_{it}\tilde{w}_{it}^\prime\theta+\sum_{t=1}^T\bar{y}_{it}\bar{w}_{it}^\prime\theta\right)\left(\sum_{t=1}^T\tilde{y}_{it}\tilde{w}_{it}-\sum_{t=1}^T\bar{y}_{it}\bar{w}_{it}\right)\left(\sum_{t=1}^T\tilde{y}_{it}\tilde{w}_{it}-\sum_{t=1}^T\bar{y}_{it}\bar{w}_{it}\right)^\prime}{\left(\sum_{\tilde{d}_i\in \mathcal{D}_{s_i}}\exp\left(\sum_{t=1}^T\tilde{y}_{it}\tilde{w}_{it}^\prime\theta\right)\right)^2}
\end{equation*}
is negative semi-definite. Therefore, $\ln \phi_\theta\left(d_i\:|\:s_i\right)$ is concave. 

Suppose that, conditional on $\mathbf{S}_i=\tilde{s}_i$, there exist at least two sample paths such that $\sum_{t=1}^T\tilde{y}_{it}\tilde{w}_{it}\neq\sum_{t=1}^T\bar{y}_{it}\bar{w}_{it}$. The second derivative is negative definite and $\ln \phi_\theta\left(d_i\:|\:\tilde{s}_i\right)$ is strictly concave. Therefore, $\ln \phi_\theta\left(d_i\:|\:\tilde{s}_i\right)$ has a unique maximizer $\hat{\theta}$. Assume that there exists an individual $i$ whose value of sufficient statistics is $\mathbf{S}_i=s_i$ such that condition (\ref{eq:identifying_stat}) holds. Under Assumption \ref{as:random_sampling}, the corresponding estimator of the sample objective function of the CMLE,
$$
\hat{Q}_N\left(\theta\right)=\frac{1}{N}\sum_{i=1}^N\ln\phi_\theta\left(d_i\:|\:s_i\right),
$$
is unique since $\hat{Q}_N$ is strictly concave.

\subsection*{Uniqueness of the Maximizer of the Population Objective Function of the CMLE}
Let $W_{it}$ and $w_{it}$ denote the corresponding regressors for the identified parameters and realized values of the regressors, respectively, with $W_{it}=\left(Y_{it-1},X_{it}\right)^\prime$. Denote the log likelihood conditional on $\mathbf{S}_i=s_i$ as $\ln \phi_\theta\left(D_i\:|\:s_i\right)$.
Following the proof of the almost sure convergence of the CMLE in \textcite{Anderson_1970}, consider 
$\ln \phi_\theta\left(D_i\:|\:s_i\right)-\ln \phi_{\theta_0}\left(D_i\:|\:s_i\right)$
 where $\theta_0$ is the true value of $\theta$. By Jensen's inequality, for any $s_i\in\mathcal{S}$,
\begin{equation*}
E_{\theta_0}\Big[\ln \phi_\theta\left(D_i\:|\:s_i\right)-\ln \phi_{\theta_0}\left(D_i\:|\:s_i\right)\:\Big|\:\mathbf{S}_i=s_i\Big]\leq \ln E_{\theta_0}\Big[\frac{\phi_\theta\left(D_i\:|\:s_i\right)}{\phi_{\theta_0}\left(D_i\:|\:s_i\right)}\:\Big|\:\mathbf{S}_i=s_i\Big]=0.
\end{equation*}
Suppose that there exist $\tilde{s}\in\mathcal{S}$ such that condition (\ref{eq:identifying_stat}) holds. Then, the inequality implies for some $\tilde{s}\in\mathcal{S}$,
$$
E_{\theta_0}\Big[\ln \phi_\theta\left(D_i\:|\:\tilde{s}\right)-\ln \phi_{\theta_0}\left(D_i\:|\:\tilde{s}\right)\:\Big|\:\mathbf{S}_i=\tilde{s}\Big]<0
$$
holds. Since there is always at least one sample path in $\mathcal{D}_s$ for all $s\in\mathcal{S}$, $g_{\theta_0}(s\:|\:\alpha_i,\mathbf{G}_i)>0$ for all $s\in\mathcal{S}$. Therefore,
$$
E_{\theta_0,\alpha_i,\mathbf{G}_i}\Big[\ln \phi_{\theta}\left(D_i\:|\:\mathbf{S}_i\right)-\ln \phi_{\theta_0} \left(D_i\:|\:\mathbf{S}_i\right)\Big]<0, \quad \forall i=1,\hdots,N. 
$$
where the expectation is taken with respect to the density $g_{\theta_0}(s_i\:|\:\alpha_i,\mathbf{G}_i)$. Under Assumption \ref{as:iida},
\begin{align*}
&E_{\theta_0,\mathbf{G}_i}\Big[\ln \phi_{\theta}\left(D_i\:|\:\mathbf{S}_i\right)-\ln \phi_{\theta_0}\left(D_i\:|\:\mathbf{S}_i\right)\Big]\\
&=\int_{\alpha}E_{\theta_0,\alpha,\mathbf{G}_i}\Big[\ln \phi_{\theta}\left(D_i\:|\:\mathbf{S}_i\right)-\ln \phi_{\theta_0}\left(D_i\:|\:\mathbf{S}_i\right)\Big]H_{\iota_i}\left(\alpha\right)d\alpha<0, \quad \forall i=1,\hdots,N. 
\end{align*}
Thus, $\theta_0$ is the unique maximizer of the expected log conditional likelihood almost surely, 
\begin{align*}
&E_{\theta_0,\mathbf{G}_i}\Big[\ln \phi_{\theta}\left(D_i\:|\:\mathbf{S}_i\right)-\ln \phi_{\theta_0}\left(D_i\:|\:\mathbf{S}_i\right)\Big]<0,  \quad \forall i=1,\hdots,N.
\end{align*}
The corresponding objective function of CMLE is
\begin{align}
Q_N\left(\theta_0\right)=\frac{1}{N}\sum_{i=1}^N E_{\theta_0,\mathbf{G}_i}\Big[\ln \phi_{\theta}\left(D_i\:|\:\mathbf{S}_i\right)-\ln \phi_{\theta_0} \left(D_i\:|\:\mathbf{S}_i\right)\Big]<0\label{ap_eq:negative_object}
\end{align}
by Assumption \ref{as:random_sampling}. Thus, the population objective function has a unique maximizer at $\theta_0$ almost surely.

\section{Proofs of Propositions and Lemmas}\label{ap:proof}

\PropnonidentificationCMLE*

\begin{proof}
Consider sufficient statistics for the feedback process, $\mathbf{G}_i$, and the unobserved heterogeneity, $\alpha_i$:
$$
\mathbf{S}^1=\left(\sum_{t=1}^{T} Y_{it}, \left(\sum_{t=1}^{T-1} \mathbbm{1}\{X_{it}=x_1,Y_{it}=y,X_{it+1}=x_2\}\right)_{x_1,x_2\in\mathcal{X},y\in\{0,1\}} \right).
$$
Without loss of generality, choose arbitrary values of the initial condition $\left(Y_{i0},X_{i1}\right)=\left(y_{0},\tilde{x}_{1}\right)$ and sufficient statistics $\mathbf{S}^1=s$,
\begin{align*}
s=\left(n_y, \left(n_{x_1,x_2,y}\right)_{x_1,x_2\in\mathcal{X},y\in\{0,1\}} \right).
\end{align*}
Note that 
$\sum_{x_1,x_2\in\mathcal{X},y\in\{0,1\}} n_{x_1,x_2,y}=T-1$ because there is $T-1$ number of transitions of $\left(X_{it},Y_{it},X_{it+1}\right)$.

First, consider the identification of $\beta$. If the sufficient statistics, $\sum_{t=1}^{T-1} \mathbbm{1}\{X_{it}=x_1,Y_{it}=y,X_{it+1}=x_2\}$, are fixed, the following quantities, which are functions of the sufficient statistics, are also determined
\begin{align}
\sum_{t=1}^{T-1}X_{it}Y_{it}&=\sum_{x_1,x_2\in\mathcal{X},y\in\{0,1\}} n_{x_1,x_2,y}\cdot x_1 y \label{ap_eq:prop1_XY_T_1}\\
\sum_{t=1}^{T-1}Y_{it}&=\sum_{x_1,x_2\in\mathcal{X},y\in\{0,1\}} n_{x_1,x_2,y}\cdot y \label{ap_eq:prop1_Y_T_1}.
\end{align}
Sum of $X_{it}$ can be represented as
$
\sum_{t=1}^{T}X_{it}=\sum_{t=1}^{T-1}X_{it}+X_{iT}=X_{i1}+\sum_{t=1}^{T-1}X_{it+1}.
$
Note that 
$$
\sum_{t=1}^{T-1}X_{it}+X_{iT}=\sum_{x_1,x_2\in\mathcal{X},y\in\{0,1\}} n_{x_1,x_2,y}\cdot x_1 +X_{iT}
$$
and 
$$
X_{i1}+\sum_{t=1}^{T-1}X_{it+1}=\tilde{x}_1+\sum_{x_1,x_2\in\mathcal{X},y\in\{0,1\}} n_{x_1,x_2,y}\cdot x_2.
$$
Therefore, $X_{iT}$ is determined as
\begin{equation}
X_{iT}=\tilde{x}_1+\sum_{x_1,x_2\in\mathcal{X},y\in\{0,1\}} n_{x_1,x_2,y}\cdot \left(x_2-x_1\right)\label{ap_eq:prop1_X_T}.
\end{equation}

Since $\sum_{t=1}^{T} Y_{it}=n_y$, it follows that
\begin{equation}
Y_{iT}=\sum_{t=1}^{T} Y_{it}-\sum_{t=1}^{T-1} Y_{it}=n_y-\sum_{x_1,x_2\in\mathcal{X},y\in\{0,1\}} n_{x_1,x_2,y}\cdot y\label{ap_eq:prop1_Y_T}.
\end{equation}
Therefore, combining equation (\ref{ap_eq:prop1_XY_T_1}), (\ref{ap_eq:prop1_X_T}), and (\ref{ap_eq:prop1_Y_T}), $\sum_{t=1}^{T}X_{it}Y_{it}$ are fixed as
\begin{multline*}
\sum_{t=1}^{T}X_{it}Y_{it}=\sum_{t=1}^{T-1}X_{it}Y_{it}+X_{iT}Y_{iT}=\sum_{x_1,x_2\in\mathcal{X},y\in\{0,1\}} n_{x_1,x_2,y}\cdot x_1 y\\
+ \left(\tilde{x}_1+\sum_{x_1,x_2\in\mathcal{X},y\in\{0,1\}} n_{x_1,x_2,y}\cdot \left(x_2-x_1\right)\right)\cdot\left(n_y-\sum_{x_1,x_2\in\mathcal{X},y\in\{0,1\}} n_{x_1,x_2,y}\cdot y\right).
\end{multline*}
Since $s$ is arbitrary, every possible sample path conditional on $\mathbf{S}^1$ has identical $\sum_{t=1}^{T}X_{it}Y_{it}$. The distribution of $\sum_{t=1}^{T}X_{it}Y_{it}$ conditional on $\mathbf{S}^1$ is degenerate. Identification of $\beta_0$ fails regardless of the time period $T$ according to condition (\ref{eq:identification_condition_logit}).

Second, consider the identification of $\rho$. From equation (\ref{ap_eq:prop1_Y_T_1}), (\ref{ap_eq:prop1_Y_T}) and the initial condition, notice that all possible path probabilities have identical 
\begin{equation}
\left(Y_{i0}, \sum_{t=1}^{T-1}Y_{it}, Y_{iT}\right)=\left(y_{0}, \sum_{x_1,x_2\in\mathcal{X},y\in\{0,1\}} n_{x_1,x_2,y}\cdot y, n_y-\sum_{x_1,x_2\in\mathcal{X},y\in\{0,1\}} n_{x_1,x_2,y}\cdot y\right).\label{ap_eq:prop1_set_Y}
\end{equation}
When $T=2$, identification of $\rho$ fails because equation (\ref{ap_eq:prop1_set_Y}) is
$$
\left(Y_{i0},Y_{i1},Y_{i2}\right)=\left(y_{0}, \sum_{x_1,x_2\in\mathcal{X},y\in\{0,1\}} n_{x_1,x_2,y}\cdot y, n_y-\sum_{x_1,x_2\in\mathcal{X},y\in\{0,1\}} n_{x_1,x_2,y}\cdot y\right)
$$
which indicates that it does not allows for any heterogeneous sequences of $\left(Y_{i0},Y_{i1},Y_{i2}\right)$. $\sum_{t=1}^T Y_{it-1}Y_{it}$ is determined as
$$
\sum_{t=1}^T Y_{it-1}Y_{it}=y_0\cdot\sum_{x_1,x_2\in\mathcal{X},y\in\{0,1\}} n_{x_1,x_2,y}\cdot y +\left(\sum_{x_1,x_2\in\mathcal{X},y\in\{0,1\}} n_{x_1,x_2,y}\cdot y\right)\cdot \left(n_y-\sum_{x_1,x_2\in\mathcal{X},y\in\{0,1\}} n_{x_1,x_2,y}\cdot y\right).
$$
for all possible $\left(X_{i}^{1:T},Y_{i}^{0:T}\right)$ conditional on $\mathbf{S}^1$. Consequently, identification of $\rho$ fails when $T=2$ according to condition (\ref{eq:identifying_stat}).

For $T\geq3$, there can be multiple possible values of $\sum_{t=1}^T Y_{it-1}Y_{it}$ since $\sum_{t=1}^{T-1}Y_{it}$ allows multiple sample paths. For example, when $T=3$, $Y_{i1}+Y_{i2}$ is 
$$
Y_{i1}+Y_{i2}=\sum_{x_1,x_2\in\mathcal{X},y\in\{0,1\}} n_{x_1,x_2,y}\cdot y
$$
Since $\sum_{x_1,x_2\in\mathcal{X},y\in\{0,1\}} n_{x_1,x_2,y}=2$ and $y$ is zero or one, $Y_{i1}+Y_{i2}$ can take values $0$,$1$, or $2$:
$Y_{i1}+Y_{i2}=0,1,2.$
If $Y_{i1}+Y_{i2}$ is zero or two, $Y_{i1}$ and $Y_{i2}$ are zero and one, respectively. There is no variation in $\sum_{t=1}^T Y_{it-1}Y_{it}$ in these cases. Thus, choose the values of the sufficient statistics where $\sum_{x_1,x_2\in\mathcal{X},y\in\{0,1\}} n_{x_1,x_2,y}\cdot y=1$. Sum of $Y_{it}$ and $Y_{i3}$ is 
$$
n_y=\sum_{t=1}^{T}Y_{it}=Y_{i1}+Y_{i2}+Y_{i3}=1+Y_{i3},\quad Y_{i3}=n_y-1.
$$
Note that $n_y$ can take values $1$ or $2$ in this case.

There are two possible sample paths which satisfy equation (\ref{ap_eq:prop1_set_Y}):
\begin{align*}
Y_{i}^{A,0:3}=\left(Y_{i0}^A,Y_{i1}^A,Y_{i2}^A,Y_{i3}^A\right)&=\left(y_{0},1,0, n_y-1\right)\\
Y_{i}^{B,0:3}=\left(Y_{i0}^B,Y_{i1}^B,Y_{i2}^B,Y_{i3}^B\right)&=\left(y_{0},0,1,n_y-1\right).
\end{align*}
Each path probability has
\begin{align*}
    \sum_{t=1}^3 Y_{it-1}^A Y_{it}^A= y_0,\quad 
    \sum_{t=1}^3 Y_{it-1}^B Y_{it}^B= n_y-1.
\end{align*}
If $y_0\neq n_y-1$, there are variations in $\sum_{t=1}^TY_{it-1}Y_{it}$ conditional on the sufficient statistics. If
$\left(y_0,n_y\right)=\left(0,2\right)$ or $\left(1,1\right)$, $\sum_{t=1}^3 Y_{it-1}^A Y_{it}^A\neq \sum_{t=1}^3 Y_{it-1}^B Y_{it}^B$. By Assumptions \ref{as:iida} - \ref{as:markov_setup1}, each path can be attainable with positive probability given the initial condition. Thus, there exist two sample paths that have different values in $\sum_{t=1}^TY_{it-1}Y_{it}$ with positive probability.  By the identification condition (\ref{eq:identifying_stat}), identification of $\rho$ is possible when time period $T=3$. The exact number of possible paths of $\left(X_{i}^{2:T},Y_{i}^{1:T}\right)$ conditional on the sufficient statistics and the initial condition depends on the cardinality of the supports, $|\mathcal{X}|$. It follows that identification of $\rho$ is possible when $T\geq3$.
\end{proof}

\Lemidenticalsumy*
\begin{proof}
Denote $\sum_{t=1}^T x_ty_t$ for the path $s\in\mathcal{D}_s$ by $\left(\sum_{t=1}^T x_ty_t\right)_s$. Note that $\sum_{x_1,x_2,y\in\{0,1\}} n_{x_1,x_2,y}^{s}=T-1$ and $\sum_{x,y\in\{0,1\}}n_{x,1,y}^{s},\:\sum_{x,y\in\{0,1\}}n_{x,0,y}^{s}\leq T-1$ since there are $T-1$ feedback processes of $X_{it}$ over $T$ period. In terms of the polynomial structure of the probability in $A_i$, the linear combination of probabilities can be represented as
\begin{multline*}
    \sum_{s\in\mathcal{S}} m_s p_s\left(\alpha_k\right)=\Big\{1+\exp\left(\beta x_1\right)A_i\Big\}^{-1}\Big\{1+\exp{\left(\beta\right)A_i\Big\}^{-T}\Big\{1+A_i}\Big\}^{-T}\sum_{s\in\mathcal{S}} m_s\exp{\left(\beta \left(\sum_{t=1}^T x_ty_t\right)_s\right)}A_i^{n_y^s}\\
\times 
\underbrace{\Big\{1+\exp{\left(\beta\right)A_i\Big\}^{T-\sum_{x,y\in\{0,1\}}n_{x,1,y}^{s}}\Big\{1+A_i}\Big\}^{T-\sum_{x,y\in\{0,1\}}n_{x,0,y}^{s}}}_{\text{polynomial of degree $T+1$ in $A_i$}}\prod_{x,y\in\{0,1\}}
G_{xy}^{n_{x,1,y}^{s}}\left(\alpha_k\right)\left(1-G_{xy}\left(\alpha_k\right)\right)^{n_{x,0,y}^{s}}=0.
\end{multline*}
The degree in terms of $A_i$ of the path $s$ is determined by $n_y^s$ because 
$$
\sum_{x,y\in\{0,1\}}n_{x,1,y}^{s}+\sum_{x,y\in\{0,1\}}n_{x,0,y}^{s}=T-1,
$$
implying the degree of
$$
\Big\{1+\exp{\left(\beta\right)A_i\Big\}^{T-\sum_{x,y\in\{0,1\}}n_{x,1,y}^{s}}\Big\{1+A_i}\Big\}^{T-\sum_{x,y\in\{0,1\}}n_{x,0,y}^{s}}
$$
is always $T+1$ in $A_i$. 

When $n_y^s$ differs across the path probabilities, the linear combination separates into groups indexed by $n_y^s$. Therefore, to match the polynomial degree in $A_i$, $n_y^s$ should be identical across nontrivially linearly dependent paths. If not, for some $s$, $m_s$ is a non-constant function of $A_i$.
\end{proof}

\LemidenticalorderG*
\begin{proof}
Similar to Lemma \ref{lm:identical_sum_y}, in terms of the polynomial structure of the probabilities in $G_{xy}$, the probabilities can be represented as
\begin{multline*}
    \sum_{s\in\mathcal{S}} m_s p_s\left(\alpha_k\right)=\Big\{1+\exp\left(\beta x_1\right)A_i\Big\}^{-1}\Big\{1+\exp{\left(\beta\right)A_i\Big\}^{-T}\Big\{1+A_i}\Big\}^{-T}\sum_{s\in\mathcal{S}} m_s \exp{\left(\beta \left(\sum_{t=1}^T x_{t}y_{t}\right)_{s}\right)}\\
\times A_i^{n_y^s}
\Big\{1+\exp{\left(\beta\right)A_i\Big\}^{T-\sum_{x,y\in\{0,1\}}n_{x,1,y}^{s}}\Big\{1+A_i}\Big\}^{T-\sum_{x,y\in\{0,1\}}n_{x,0,y}^{s}}\\
\prod_{x,y\in\{0,1\}}
\underbrace{G_{xy}^{n_{x,1,y}^{s}}\left(\alpha_k\right)\left(1-G_{xy}\left(\alpha_k\right)\right)^{n_{x,0,y}^{s}}}_{\text{polynomial of degree $n_{x,y}^s=n_{x,0,y}^{s}+n_{x,1,y}^{s}$ in $G_{xy}\left(\alpha_k\right)$}}=0
\end{multline*}
If the probabilities of the paths are non-trivially linearly dependent, the non-trivial dependence must be confined to paths with identical polynomial degrees in $G_{xy}\left(\alpha_k\right)$. Therefore, the paths have identical values of $n_{x,y}$ where $n_{x,y}=n_{x,y}^s=n_{x,1,y}^{s}+n_{x,0,y}^{s}$ for all $x,y\in\{0,1\}$. If not, for some $s$, $m_s$ is a non-constant function of $\mathbf{G}$. 
\end{proof}

\Lemnonotrivial*
\begin{proof}
Under Lemma \ref{lm:identical_sum_y} and Lemma \ref{lm:identical_order_G}, if the path probabilities have nontrivial linear dependence, the probabilities have identical values of the statistics
$$
\sum_{t=1}^T Y_{it}, \quad \sum_{t=2}^T\mathbbm{1}\{X_{it-1}=x_1,Y_{it-1}=y\}.
$$
Since Lemma \ref{lm:identical_order_G} fixes $n_{x,y}=n_{x,y}^s=n_{x,1,y}^{s}+n_{x,0,y}^{s}$ for all $x,y\in\{0,1\}$, any remaining variation must occur within such pairs, and it is sufficient to consider one varying pairs. Suppose the linearly dependent paths have identical values of the statistics given by
$$
n_y^s=n_y,\quad n_{0,1,0}^{s}+n_{0,0,0}^{s}=n_{0,0},\quad n_{0,1,1}^{s}+n_{0,0,1}^{s}=n_{0,1},\quad n_{1,1,0}^{s}+n_{1,0,0}^{s}=n_{1,0},\quad n_{1,1,1}^{s}+n_{1,0,1}^{s}=n_{1,1}.
$$
Also, assume further that each value of $n_{x_1,x_2,y}^{s}$ among the paths is identical except for $n_{0,0,0}^{s}$ and $n_{0,1,0}^{s}$.
$$
n_{x_1,x_2,y}^{s}=n_{x_1,x_2,y}, \quad\forall  \left(x_1,x_2,y\right)\in\{0,1\}^{3}\setminus \{\left(0,0,0\right),\left(0,1,0\right)\}
$$
The linearly dependent path probabilities differ only in $n_{0,0,0}^{s}$ and $n_{0,1,0}^{s}$, so they can be written as
\begin{multline*}
    \sum_{s\in\mathcal{S}} m_s p_s\left(\alpha_k\right)=\Big\{1+\exp\left(\beta x_1\right)A_i\Big\}^{-1}A_i^{n_{y}} \Big\{1+\exp{\left(\beta\right)A_i\Big\}^{-n_{0,1,1}-n_{1,1,0}-n_{1,1,1}}\Big\{1+A_i}\Big\}^{-n_{0,0,1}-n_{1,0,0}-n_{1,0,1}}\\
\times G_{01}^{n_{0,1,1}}\left(\alpha_k\right)\left(1-G_{01}\left(\alpha_k\right)\right)^{n_{0,0,1}}G_{10}^{n_{1,1,0}}\left(\alpha_k\right)\left(1-G_{10}\left(\alpha_k\right)\right)^{n_{1,0,0}}G_{11}^{n_{1,1,1}}\left(\alpha_k\right)\left(1-G_{11}\left(\alpha_k\right)\right)^{n_{1,0,1}}
    \\
\times \sum_{s\in\mathcal{S}} \underbrace{m_s\exp{\left(\left(\beta \sum_{t=1}^T x_{t}y_{t}\right)_s\right)}
\Big\{1+\exp{\left(\beta\right)A_i\Big\}^{-n_{0,1,0}^{s}}\Big\{1+A_i}\Big\}^{-n_{0,0,0}^{s}}}_{=\tilde{m}_s}
G_{00}^{n_{0,1,0}^{s}}\left(\alpha_k\right)\left(1-G_{00}\left(\alpha_k\right)\right)^{n_{0,0,0}^{s}}=0
\end{multline*}
Therefore, it is equivalent to identifying a linear relation among $G_{00}^{n_{0,1,0}^{s}}\left(\alpha_k\right)\left(1-G_{00}\left(\alpha_k\right)\right)^{n_{0,0,0}^{s}}$ for all $s\in\mathcal{S}$,
$$
\sum_{s\in\mathcal{S}} \tilde{m}_s
G_{00}^{n_{0,1,0}^{s}}\left(\alpha_k\right)\left(1-G_{00}\left(\alpha_k\right)\right)^{n_{0,0,0}^{s}}=\sum_{s\in\mathcal{S}} \tilde{m}_s
G_{00}^{n_{0,1,0}^{s}}\left(\alpha_k\right)\left(1-G_{00}\left(\alpha_k\right)\right)^{n_{0,0}-n_{0,1,0}^{s}}=0.
$$
Notice that 
$G_{00}^{n_{0,1,0}^{s}}\left(\alpha_k\right)\left(1-G_{00}\left(\alpha_k\right)\right)^{n_{0,0}-n_{0,1,0}^{s}}$
are Bernstein basis polynomials of degree $n_{0,0}$ in $G_{00}\left(\alpha_k\right)$ which are linearly independent. Therefore, 
$$
\tilde{m}_s=m_s\exp{\left(\beta \left(\sum_{t=1}^T x_{t}y_{t}\right)_s\right)}
\Big\{1+\exp{\left(\beta\right)A_i\Big\}^{-n_{0,1,0}^{s}}\Big\{1+A_i}\Big\}^{-n_{0,0,0}^{s}}=0, \quad \forall s \in\mathcal{S}.
$$
Since $\exp{\left(\left(\beta \sum_{t=1}^T x_{t}y_{t}\right)_s\right)}
\Big\{1+\exp{\left(\beta\right)A_i\Big\}^{-n_{0,1,0}^{s}}\Big\{1+A_i}\Big\}^{-n_{0,0,0}^{s}}$ is positive, it implies that
$$
m_s=0, \quad \forall s \in\mathcal{S}.
$$
Thus, there are no sample paths whose probabilities are non-trivially linearly dependent.
\end{proof}

\propsetupyonly*
\begin{proof}
Consider following sufficient statistics for $\mathbf{G}_i$, and $\alpha_i$:
$$
\mathbf{S}^2=\left(\sum_{t=1}^T Y_{it}, \left(\sum_{t=1}^{T} \mathbbm{1}\{Y_{it-1}=y,X_{it}=x\}\right)_{y\in\{0,1\},x\in\mathcal{X}}\right).
$$
Similar to the proof of Proposition \ref{pp:setup1}, choose arbitrary values of the initial condition $Y_{i0}=y_{0}$ and the sufficient statistics $\mathbf{S}^2=s$,
\begin{align*}
s=\left(n_y, \left(n_{y,x}\right)_{y\in\{0,1\},x\in\mathcal{X}} \right).
\end{align*}
Since there are $T$ number of transitions of $\left(Y_{it-1},X_{it}\right)$, $\sum_{y\in\{0,1\},x\in\mathcal{X}} n_{y,x}=T$.

First, consider the identification of $\beta$. From 
\begin{align*}
    \sum_{t=1}^T X_{it}&=\sum_{y\in\{0,1\},x\in\mathcal{X}} n_{y,x}\cdot x,\quad \sum_{t=1}^{T} Y_{it-1}=\sum_{y\in\{0,1\},x\in\mathcal{X}} n_{y,x}\cdot y\\
    Y_{iT}&=\sum_{t=1}^{T} Y_{it}-\sum_{t=1}^{T-1} Y_{it}=\sum_{t=1}^{T} Y_{it}-\left(\sum_{t=1}^{T} Y_{it-1}-Y_{i0}\right)=n_y-\sum_{y\in\{0,1\},x\in\mathcal{X}} n_{y,x}\cdot y+y_0
\end{align*}
The statistics $\mathbf{S}^2$ do not constrain $X_{iT}$ to be identical, which is less restrictive than the specification under Assumption \ref{as:markov_setup1}. Conditional on $\mathbf{S}^2=s$, all possible $\left(X_{i}^{2:T},Y_{i}^{1:T}\right)$ have identical
\begin{equation*}
\left(Y_{i0},\sum_{t=1}^{T-1}Y_{it},Y_{iT}\right)=\left(y_{0},\sum_{y\in\{0,1\},x\in\mathcal{X}} n_{y,x}\cdot y-y_0,n_y-\sum_{y\in\{0,1\},x\in\mathcal{X}} n_{y,x}\cdot y+y_0\right).\label{ap_eq:prop2_set_Y}
\end{equation*}
When $T=2$, it is equivalent to $\left(Y_{i0},\sum_{t=1}^{T-1}Y_{it},Y_{iT}\right)=\left(Y_{i0},Y_{i1},Y_{i2}\right)$. 

Observe that given sufficient statistics, there exist at most two sample paths of $\left(X_{i}^{1:2},Y_{i}^{0:2}\right)$ since all path share identical $Y_{i0},Y_{i2}$ and $\left(\sum_{t=1}^{T} \mathbbm{1}\{Y_{it-1}=y,X_{it}=x\}\right)_{y\in\{0,1\},x\in\mathcal{X}}$.
All paths have the transition of $X_{i1}$ given identical initial condition, $Y_{i0}=y_0$. Without loss of generality, for arbitrary $x_1,x_2\in\mathcal{X}$, let $n_{y_0,x_1}=1$ and $n_{y_0,x_2}=1$ be the transition of $X_{i1}$ given $y_{0}$ in each path. Since the paths hold an identical number of transitions of $X_{it}$ given $Y_{it-1}$,
$$
\sum_{y\in\{0,1\},x\in\mathcal{X}} n_{y,x}\cdot y=n_{y_0,x_1}\cdot y_0+n_{y_0,x_2}\cdot y_0=2y_0
$$
holds for every paths. Therefore, the path $\left(Y_{i0},Y_{i1},Y_{i2}\right)$ is given as
\begin{equation}
\left(Y_{i0},Y_{i1},Y_{i2}\right)=\left(y_{0},\sum_{y\in\{0,1\},x\in\mathcal{X}} n_{y,x}\cdot y-y_0,n_y-\sum_{y\in\{0,1\},x\in\mathcal{X}} n_{y,x}\cdot y+y_0\right)=\left(y_0,y_0,n_y-y_0\right).\label{ap_eq:prop2_set_Y_t2}
\end{equation}
Two possible paths are thus
$$
\begin{array}{cccc}
&Y_{i0}&(X_{i1},Y_{i1}) &(X_{i2},Y_{i2})\\\hline\hline
A:&y_0 & (x_1,y_0) &(x_2,n_y-y_0) \\
B:&y_0& (x_2,y_0) &(x_1,n_y-y_0)\\
\end{array}
$$
where each path has
\begin{align*}
    \sum_{t=1}^2 X_{it}^A Y_{it}^A= x_1y_0+x_2\left( n_y-y_0\right),\quad 
    \sum_{t=1}^2 X_{it}^B Y_{it}^B= x_2y_0+x_1\left(n_y-y_0\right).
\end{align*}
To hold $\sum_{t=1}^2 X_{it}^A Y_{it}^A\neq\sum_{t=1}^2 X_{it}^B Y_{it}^B$ to satisfy condition (\ref{eq:identifying_stat}),
$$
\left(x_1-x_2\right)\left(2y_0-n_y\right)\neq0
$$
If $x_1\neq x_2$ and $2y_0\neq n_y$, there are variations in $\sum_{t=1}^2 X_{it} Y_{it}$. When $y_0=0$, if $Y_{i2}=1$ $n_y=Y_{i1}+Y_{i2}=1$, $2y_0\neq n_y$. Likewise, when $y_0=1$, if $Y_{i2}=0$, $n_y=Y_{i1}+Y_{i2}=1$ and $2y_0\neq n_y$. Therefore, for any $x_1,x_2\in\mathcal{X}$ and $x_1\neq x_2$, following path probabilities have variations in $\sum_{t=1}^T X_{it}Y_{it}$:
$$
\begin{array}{cccc}
&Y_{i0}&(X_{i1},Y_{i1}) &(X_{i2},Y_{i2})\\\hline\hline
A:&0 & (x_1,0) &(x_2,1) \\
B:&0& (x_2,0) &(x_1,1)\\
\end{array},
\quad
\begin{array}{cccc}
&Y_{i0}&(X_{i1},Y_{i1}) &(X_{i2},Y_{i2})\\\hline\hline
A:&1 & (x_1,1) &(x_2,0) \\
B:&1& (x_2,1) &(x_1,0)\\
\end{array}.
$$
Therefore, $\beta$ is identified when $T=2$. It follows that $\beta$ is identified with $T\ge3$.

Identification of $\rho$ is not plausible when $T=2$ because $\left(Y_{i0},Y_{i1},Y_{i2}\right)$ is determined as described in (\ref{ap_eq:prop2_set_Y_t2}). In this case, $\sum_{t=1}^TY_{it-1}Y_{it}$ is 
$$
\sum_{t=1}^TY_{it-1}Y_{it}=y_0^2+y_0\left(n_y-y_0\right).
$$
Therefore, there is no variation in $\sum_{t=1}^TY_{it-1}Y_{it}$ when $T=2$.


It is possible to identify $\rho$ when $T\geq3$. $\mathbf{S}^2$ do not impose more constraints on possible $\left(X_{i}^{1:T},Y_{i}^{1:T}\right)$ than those of $\mathbf{S}^1$ from the proof of Proposition \ref{pp:setup1} in a sense that $\mathbf{S}^2$ and initial conditin $Y_{i0}$ allow variations in $X_{i1},X_{iT}$ and impose weaker constraints on the transition of $X_{it}$. Thus, identification of $\rho$ is equivalent to identification of $\rho$ in the previous arguments with more possible set of the paths. Therefore, identification of $\rho$ is possible when $T\geq3$.
\end{proof}

\propsetupinitial*

\begin{proof}
Consider sufficient statistics for the feedback process, $\mathbf{G}_i$, and the unobserved heterogeneity, $\alpha_i$:
$$
\mathbf{S}^3=\left(\sum_{t=1}^{T} Y_{it}, \left(\sum_{t=1}^{T} \mathbbm{1}\{X_{it-1}=x_1,Y_{it-1}=y,X_{it}=x_2\}\right)_{x_1,x_2\in\mathcal{X},y\in\{0,1\}} \right).
$$
Similar to the proof of Proposition \ref{pp:setup1} and \ref{pp:setup2}, let the arbitrary values of the sufficient statistics $\mathbf{S}^3=s$ be
\begin{align*}
s=\left(n_y, \left(n_{x_1,x_2,y}\right)_{x_1,x_2\in\mathcal{X},y\in\{0,1\}} \right).
\end{align*}
Notice that the initial condition $\left(X_{i0},Y_{i0}\right)$ of the paths does not need to be same, which provides the source of identification. 

First, consider the identification of $\beta$. The following quantities are determined as the sufficient statistics are fixed,
\begin{align*}
\sum_{t=0}^{T-1}X_{it}Y_{it}=\sum_{x_1,x_2\in\mathcal{X},y\in\{0,1\}} n_{x_1,x_2,y}\cdot x_1 y,\quad
\sum_{t=0}^{T-1}Y_{it}=\sum_{x_1,x_2\in\mathcal{X},y\in\{0,1\}} n_{x_1,x_2,y}\cdot y.
\end{align*}
Sum of $X_{it}$ is
$
\sum_{t=0}^{T}X_{it}=\sum_{t=0}^{T-1}X_{it}+X_{iT}=X_{i1}+\sum_{t=0}^{T-1}X_{it+1}
$
where left hand side of the sum is
$$
\sum_{t=0}^{T-1}X_{it}+X_{iT}=\sum_{x_1,x_2\in\mathcal{X},y\in\{0,1\}} n_{x_1,x_2,y}\cdot x_1 +X_{iT}
$$
and right hand side of the sum is
$$
X_{i0}+\sum_{t=0}^{T-1}X_{it+1}=X_{i0}+\sum_{x_1,x_2\in\mathcal{X},y\in\{0,1\}} n_{x_1,x_2,y}\cdot x_2.
$$
Unlike the result of Proposition \ref{pp:setup1}, $X_{i1}$ is not an initial condition. Thus, $X_{iT}$ can be varied with $X_{i1}$
\begin{equation*}
X_{iT}-X_{i0}=\sum_{x_1,x_2\in\mathcal{X},y\in\{0,1\}} n_{x_1,x_2,y}\cdot \left(x_2-x_1\right).
\end{equation*}

Since $\sum_{t=1}^{T} Y_{it}=n_y$, it follows that
\begin{equation}
Y_{iT}=\sum_{t=1}^{T} Y_{it}-\sum_{t=0}^{T-1} Y_{it}+Y_{i0}=n_y-\sum_{x_1,x_2\in\mathcal{X},y\in\{0,1\}} n_{x_1,x_2,y}\cdot y+Y_{i0}\label{ap_eq:prop4_Y_T}.
\end{equation}
Therefore, $\sum_{t=0}^{T}X_{it}Y_{it}$ are
\begin{multline*}
\sum_{t=0}^{T}X_{it}Y_{it}=\sum_{t=0}^{T-1}X_{it}Y_{it}+X_{iT}Y_{iT}=\sum_{x_1,x_2\in\mathcal{X},y\in\{0,1\}} n_{x_1,x_2,y}\cdot x_1 y\\
+ \left(X_{i0}+\sum_{x_1,x_2\in\mathcal{X},y\in\{0,1\}} n_{x_1,x_2,y}\cdot \left(x_2-x_1\right)\right)\cdot\left(Y_{i0}+n_y-\sum_{x_1,x_2\in\mathcal{X},y\in\{0,1\}} n_{x_1,x_2,y}\cdot y\right).
\end{multline*}
$\sum_{t=1}^{T}X_{it}Y_{it}$ is
\begin{multline*}
\sum_{t=1}^{T}X_{it}Y_{it}=\sum_{t=0}^{T}X_{it}Y_{it}-X_{i0}Y_{i0}=\sum_{x_1,x_2\in\mathcal{X},y\in\{0,1\}} n_{x_1,x_2,y}\cdot x_1 y\\
+ Y_{i0}\sum_{x_1,x_2\in\mathcal{X},y\in\{0,1\}} n_{x_1,x_2,y}\cdot \left(x_2-x_1\right)
+X_{i0}\left(n_y-\sum_{x_1,x_2\in\mathcal{X},y\in\{0,1\}} n_{x_1,x_2,y}\cdot y\right)
\\+ \left(\sum_{x_1,x_2\in\mathcal{X},y\in\{0,1\}} n_{x_1,x_2,y}\cdot \left(x_2-x_1\right)\right)\cdot\left(n_y-\sum_{x_1,x_2\in\mathcal{X},y\in\{0,1\}} n_{x_1,x_2,y}\cdot y\right).
\end{multline*}
Thus, depending on the initial condition $\left(X_{i0},Y_{i0}\right)$, there can be variations in $\sum_{t=1}^{T}X_{it}Y_{it}$.

Notice that $\sum_{t=1}^{T}X_{it}Y_{it}$ can vary depending on the initial condition $X_{i1}$ except when $T=1$. When $T=1$, there is one transition of $X_{i1}$ given $\left(X_{i0},Y_{i0}\right)$, 
$\sum_{x_1,x_2\in\mathcal{X},y\in\{0,1\}} n_{x_1,x_2,y}=1$. Since there is one feedback process of $X_{i1}$ given $\left(X_{i0},Y_{i0}\right)$, each path should have identical initial condition and $X_{i1}$ to hold identical feedback of $X_{i1}$. Assume it is determined as
\begin{equation}
\left(X_{i0},Y_{i0},X_{i1}\right)=\left(\tilde{x}_0,\tilde{y}_0,\tilde{x}_1\right).\label{ap_eq:prop4_T=2}
\end{equation}
Since $\sum_{t=1}^1Y_{it}=Y_{i1}$, each paths also has identical $Y_{i1}=n_y$. Thus,
$$
\sum_{t=1}^{1}X_{it}Y_{it}=\tilde{x}_1  n_y.
$$
The identification condition (\ref{eq:identifying_stat}) does not hold when $T=1$. Thus, $\beta$ is identified via conditional likelihood when $T\geq2$.

Now consider identification of $\rho$. When the initial condition is fixed, $\rho$ is identified for $T\geq3$ as described in Proposition \ref{pp:setup1}. Under the relaxed initial-condition assumption, $\rho$ is identified when $T\geq3$. Thus, now consider identification of $\rho$ when $T=2$. Observe there exist two sample paths of $\left(X_{i}^{0:2},Y_{i}^{0:2}\right)$ because there are only two feedback processes of $X_{i1}$ and $X_{i2}$. Notice that following two paths are only possible paths when $T=2$,
$$
\begin{array}{cccc}
&(X_{i0},Y_{i0})&(X_{i1},Y_{i1}) &(X_{i2},Y_{i2})\\\hline\hline
A:&(x_{0}^A,y_{0}^A)  & (x_{0}^B,y_{0}^B) &(x_{0}^A,n_y-y_{0}^B) \\
B:&(x_{0}^B,y_{0}^B)  & (x_{0}^A,y_{0}^A) &(x_{0}^B,n_y-y_{0}^A)\\
\end{array}.
$$
Suppose $y_{0}^A=y_{0}^B$. Then there is no variation in $\sum_{t=1}^T y_{t-1}y_{t}$. For the case of $y_{0}^A\neq y_{0}^B$, without loss of generality, assume $y_{0}^A=1$ and $y_{0}^B=0$. Then $n_y=1$ because $Y_{i2}$ is a binary variable. Therefore, following two paths are only possible paths,
$$
\begin{array}{cccc}
&(X_{i0},Y_{i0})&(X_{i1},Y_{i1}) &(X_{i2},Y_{i2})\\\hline\hline
A:&(x_{0}^A,1)  & (x_{0}^B,0) &(x_{0}^A,1) \\
B:&(x_{0}^B,0)  & (x_{0}^A,1) &(x_{0}^B,0)\\
\end{array}.
$$
Thus, there is also no variation in $\sum_{t=1}^T Y_{it-1}Y_{it}$.
 Consequently, $\rho$ is identified via conditional likelihood when $T\geq3$.
\end{proof}

\end{document}